\documentclass[12pt]{article}
\usepackage[mathscr]{eucal}
\usepackage[usenames, dvipsnames]{color}
\usepackage{epsfig,amsfonts}
\usepackage[body={170mm,247mm}]{geometry}
\usepackage{graphicx}
\usepackage[margin=15pt,font=small,labelfont=bf]{caption}
\usepackage{amsmath}
\usepackage{amsthm,amssymb}
\usepackage{graphicx}
\usepackage{hhline}
\usepackage{cite}
\usepackage{psfrag}
\usepackage{enumerate}
\usepackage{upgreek}
\makeatletter
\@addtoreset{equation}{section}
\makeatother

\newcounter{app}
\newcounter{sapp}[app]

\def\be{\begin{equation}}
\def\ee{\end{equation}}
\def\bdm{\begin{displaymath}}
\def\edm{\end{displaymath}}
\def\bea{\begin{eqnarray}}
\def\eea{\end{eqnarray}}

\def\ri{{\rm i}}
\def\Mba{{\mathcal M}}
\def\mye{{\phi}}

\def\XXint#1#2#3{{\setbox0=\hbox{$#1{#2#3}{\int}$}
    \vcenter{\hbox{$#2#3$}}\kern-.5\wd0}}

\def\hf{{\frac{1}{2}}}

\def\be{\begin{equation}}
\def\ee{\end{equation}}

\def\beq{\begin{equation}}
\def\eeq{\end{equation}}

\newcommand{\rd}{\mbox{d}}
\newcommand{\re}{\mbox{e}}

\def\ds{\displaystyle}
\newcommand{\JJ}{{\mathscr J}}

\begin{document}
\begin{titlepage}
{}
\vspace{1.3cm}
{}
\begin{flushright}
\end{flushright}
\vspace{2.3cm}

\begin{center}
\begin{LARGE}
{\bf Vacuum energy \\}
\end{LARGE}
\vspace{.7cm}
\begin{LARGE}
{\bf of the Bukhvostov-Lipatov model}
\end{LARGE}

\vspace{2.3cm}
\begin{large}

{\bf Vladimir V. Bazhanov$^{1}$, Sergei  L. Lukyanov$^{2,3}$ and  Boris A. Runov$^{1}$}

\end{large}

\vspace{1.cm}
$^1$Department of Theoretical Physics\\
         Research School of Physics and Engineering\\
    Australian National University, Canberra, ACT 2601, Australia\\\ \\
${}^{2}$NHETC, Department of Physics and Astronomy\\
     Rutgers University\\
     Piscataway, NJ 08855-0849, USA\\
\vspace{.2cm}
and\\
\vspace{.2cm}
${}^{3}$L.D. Landau Institute for Theoretical Physics\\
  Chernogolovka, 142432, Russia\\
\end{center}
\vspace{.5cm}
\begin{center}
\centerline{\bf Abstract} \vspace{.8cm}\small
\parbox{16cm}{
Bukhvostov and Lipatov have shown that weakly interacting 
instantons and anti-instantons in the $O(3)$ non-linear sigma model in
two dimensions are described by an exactly soluble  
model containing two coupled Dirac fermions. 
We propose an exact
formula for the vacuum energy of the model for twisted boundary
conditions, expressing it through a
special solution of the classical sinh-Gordon equation. The formula
perfectly matches predictions of the standard renormalized
perturbation theory at weak couplings as well as the conformal
perturbation theory at short distances. Our results also agree with
the Bethe ansatz solution of the model. A complete proof the proposed
expression for the vacuum energy based on a combination of 
the Bethe ansatz techniques and the classical inverse scattering transform  method 
is presented in the second part of
this work \cite{BLR:2016b}. 
}
\end{center}
\vspace{.8cm}

\vfill

\end{titlepage}
\newpage
\addtocounter{page}{1}

\section{\label{intro}Introduction}

The ``instanton calculus''  is  a common approach for studying the
non-perturbative  semiclassical effects in gauge
theories and sigma models. 
One of the first and perhaps the best known illustration  of this approach
is 
the $O(3)$ Non-Linear Sigma Model (NLSM) in two dimensions, where
multi-instanton configurations 
admit a simple analytic form \cite{Polyakov:1975yp}.
It is less known that the $O(3)$ NLSM provides an
opportunity to explore a mechanism of exact summation
of the  instanton configurations in the path integral.
In order to explain the purpose of this paper,
we start with  a brief overview of 
the main ideas
behind this summation.

The instanton  contributions  in the
$O(3)$ NLSM were calculated in a semiclassical approximation
in the paper \cite{Fateev:1979dc}.
It was shown that the effect of instantons with positive
topological charge
can be  described in terms of the non-interacting theory of  Dirac fermions.
Moreover, 
every instanton has its anti-instanton counterpart with the
same action and opposite topological charge.
Thus, neglecting  the instanton-anti-instanton interaction,  one
arrives to the theory with
two non-interacting  fermions.
Although 
the classical equation has no solutions  containing both 
instanton-anti-instanton configurations, such configurations 
must still be taken into
account. 
In ref.\,\cite{Bukhvostov:1980sn} Bukhvostov and Lipatov (BL) have found that
the weak instanton-anti-instanton interaction is described 
by means of a theory of
two  Dirac fermions, $\psi_\sigma \ (\sigma=\pm)$, with the Lagrangian
\bea\label{Lagr1}
{\cal L}=
\sum_{\sigma=\pm }{\bar \psi}_\sigma
\big(\ri \gamma^\mu\partial_\mu-M\big){ \psi}_\sigma-
g\, \big({\bar \psi}_+\gamma^\mu{ \psi}_+\big)
\big({\bar \psi}_- \gamma_\mu{ \psi}_-\big)\ .
\eea
The perturbative treatment of \eqref{Lagr1}
leads to ultraviolet (UV) divergences and requires renormalization.
The  renormalization can be performed by adding
the following
 counterterms to the Lagrangian which preserve
the invariance w.r.t.
two independent 
$U(1)$ rotations
$\psi_\pm\mapsto \re^{\ri\alpha_\pm} \, \psi_\pm$, as well as the permutation 
$\psi_+\leftrightarrow\psi_-$:
\bea\label{Lagr2}
{\cal L}_{\rm BL}={\cal L}-\sum_{\sigma=\pm}\Big(\,\delta M\, 
{\bar \psi}_\sigma{ \psi}_\sigma+
\frac{g_1}{2}\, \big({\bar \psi}_\sigma\gamma^\mu{ \psi}_\sigma\big)^2\Big)\ .
\eea
In fact the cancellation of the UV divergences
leaves undetermined one of the counterterm couplings.
It is possible
to use  the renormalization scheme where the
renormalized mass $M$, the bare mass $M_0=M+\delta M$ and UV cut-off energy scale
$\Lambda_{\rm UV}$ obey the
relation
\bea\label{aoisasosa}
\frac{M}{M_0}=\bigg(\frac{M}{\Lambda_{\rm UV}}\bigg)^\nu\ ,
\eea
where the  exponent $\nu$ is a  renormalization group invariant parameter
as well  as dimensionless coupling $g$.
For $\nu=0$ the fermion mass does not require renormalization and
the only divergent quantity  is  the zero point energy.
The theory, in a sense,  turns out to be UV finite in this case.
Then the specific {\it logarithmic} divergence of the zero point energy
can be interpreted as a ``small-instanton'' divergence in the
context of  $O(3)$ NLSM.
Recall, that the standard
lattice description
of the $O(3)$ sigma model has problems -- for
example, the lattice topological
susceptibility does not obey naive scaling
laws. L\"uscher  has shown \cite{Luscher:1981tq}
that this is because of the
so-called ``small instantons'' --
field configurations such as the winding of
the $O(3)$-field around  plaquettes of  lattice size,
giving rise to spurious contribution to  quantities related to
the zero point energy.

To the best of our 
knowledge, there is no any indication that
the fermionic QFT is integrable for general values of the parameters $(g,\nu)$  \cite{Ameduri:1998ah}.
However, 
it is expected to be an integrable  theory for $\nu=0$,
which is  of prime interest for the problem of instanton summation.
The corresponding factorizable scattering theory was proposed in
\cite{Fateev:1996ea}, by extending previous results of
\cite{ZZ79,Polyakov:1983tt,Faddeev:1985qu}.
According to the work  \cite{Fateev:1996ea}
the spectrum of the model contains a fundamental 
quadruplet of mass $M$ whose  two-particle $S$-matrix 
is given by  the direct product $(-S_{a_1}\otimes S_{a_2})$ of
two $U(1)$-symmetric solutions of the $S$-matrix bootstrap. Each of the factors
$S_a$ coincides with the soliton $S$-matrix in the quantum sine-Gordon theory with
the renormalized coupling constant $a$. The couplings
are not independent but satisfy the
condition $a_1+a_2=2$, so that,
without loss of generality, one can set $a_1=1-\delta$ and $a_2=1+\delta$ with 
$\delta\geq 0$.
A relation between  $\delta$ and  the 
four-fermion coupling $g$ is not universal, i.e.,  depends on
regularization procedure involved in the perturbative calculations.
Nevertheless,
$g=\frac{\pi \delta }{1-\delta^2}$  if one uses  the regularization  that 
preserves the underlying $U(1)\otimes U(1)$ symmetry of the BL model.
Together with the fundamental particles of mass $M$,
there are also bound states whose masses are given by
$M_n=2M\, \sin\big(\textstyle{\frac{\pi n  }{2}}\,(1-\delta) \big)$, where
the integer $n$ run from $1$ to an integer part of $\frac{1}{1-\delta}$.
As  $\delta\to 1^-$, the fermion coupling
$g$ approaches infinity $g\to\infty$, and
an increasing number of  particles with vanishing mass occur in the theory.
The theory 
can also be continued into the strong coupling regime with
$\delta>1$  by means of the  bosonization technique.
Namely, the fermionic BL model can be equivalently formulated as
a theory of
two Bose scalars $\varphi_i$
governed by the Lagrangian \cite{Bukhvostov:1980sn}
\bea\label{jssksjsak}
{\tilde {\cal L}}_{\rm BL}={\textstyle
\frac{1}{16\pi}}\ \big(\, (\partial_\nu\varphi_1)^2+(\partial_\nu\varphi_2)^2\,\big)
+4\mu\ \cos\big({\textstyle
\frac{\sqrt{a_1}}{2}}\varphi_1\big) \cos\big({\textstyle
\frac{\sqrt{a_2}}{2}}\varphi_2\big)\ .
\eea
The interacting term here can be written as $4\mu\, \cosh\big({\textstyle
\frac{\sqrt{\delta-1}}{2}}\varphi_1\big) \cos\big({\textstyle
\frac{\sqrt{\delta+1}}{2}}\varphi_2\big)$, and, hence,
the bosonic description  is still applicable  as
$\delta>1$.
As it was pointed out by Al.B.  Zamolodchikov  (unpublished, see also \cite{Fateev:1996ea}),
the Lagrangian \eqref{jssksjsak}
with $a_1=2-a_2<0$  provides a dual description of the so-called 
sausage model \cite{Fateev:1992tk}, which is a 
NLSM whose target space has a geometry
of a deformed 2-sphere. As $a_1\to-\infty$
the sausage metric gains the $O(3)$-invariance
and we come back to the $O(3)$ NLSM.
Notice that the formal substitution  $\delta\equiv 1-a_1=\infty$ into 
the relation $g=\frac{\pi\delta}{1-\delta^2}$
leads to the vanishing
fermionic coupling in the initial Lagrangian \eqref{Lagr1}.

Putting the theory on a finite segment
$x^{1}\in [0,R]$, one should impose boundary conditions on the
fundamental fermion fields.  We shall consider the twisted (quasiperiodic)
boundary conditions, which preserve the $U(1)\otimes U(1)$ invariance of the
bulk Lagrangian,
\bea\label{apssspps}
{\psi}_\pm(x^0,x^1+R)=-\re^{2\pi\ri k_\pm}\, {\psi}_\pm(x^0,x^1)\ ,\ \ \ \ 
{\bar \psi}_\pm(x^0,x^1+R)=-\re^{-2\pi\ri k_\pm}\, {\bar \psi}_\pm(x^0,x^1)\ .
\eea
The pair of real numbers $(k_+,k_-)$ labels different sectors of the
theory and, therefore, one can address the problem of computing of
vacuum energy $E_{\bf k}$ in each sector.  Notice that twisted
boundary conditions is of special interest for application of
resurgence theory to the problem of instanton summation
\cite{Dunne:2012ae}.

There is no doubt to say that the above scenario of the instanton
summation deserves a detailed quantitative study. Perhaps the simplest
question in this respect concerns an exact description of finite
volume energy spectrum for the theory \eqref{jssksjsak} in both
regimes $0<\delta< 1$ and $\delta>1$.  In this work we will focus on
the perturbative regime $0<\delta< 1$, where the fermionic description
\eqref{Lagr2} can be applied. 
We propose an exact formula which expresses the
vacuum energies in terms of certain solutions of the classical
sinh-Gordon equation. The formula is perfectly matching both the conformal
perturbation theory as well as the standard
renormalized perturbation theory for the Lagrangian \eqref{Lagr2}.
The result also agrees  with the 
original coordinate Bethe ansatz
solution of ref.\cite{Bukhvostov:1980sn} and the associated non-linear
integral equations derived in \cite{Saleur:1998wa}.
The aim of this paper is to review and further develop all these
approaches to facilitate future 
considerations of the NLSM  regime of the theory with $\delta>1$. 

The paper is organized as follows. In the first two sections we discuss
the perturbative approaches for calculating  $E_{\bf k}$.  In
Sec.\,\ref{sec2}, the small-$R$ behavior of $E_{\bf k}$ is studied by
means of the conformal perturbation theory for the bosonic Lagrangian
\eqref{jssksjsak}.  Then, in Sec.\,\ref{sec3}, using the fermionic
Lagrangian \eqref{Lagr2}, the vacuum energies are calculated within
the second order of standard renormalized perturbation theory. 
The exact formula for the
vacuum energies expressed through solutions of the classical
sinh-Gordon equation is presented in Sec.~\ref{secnew5}. 
Our considerations there are essentially based on the
previous works \cite{Lukyanov:2013wra,
Bazhanov:2013cua,Bazhanov:2014joa}.  
These connections allows one to derive a system non-linear integral equations
which is well suited for perturbative analysis around
$\delta=0$. Finally, Sec.\,\ref{sec5} contains a summary of the original 
coordinate Bethe ansatz results \cite{Bukhvostov:1980sn} and the
corresponding non-linear integral equations \cite{Saleur:1998wa}, as
well as their numerical comparison with our calculation.  

\section{\label{sec2}Small-$R$ expansion}

In this paper we shall mainly focus on 
the  BL
model with the vanishing exponent  $\nu$
\eqref{aoisasosa}. Nevertheless  
it is useful to start with the theory characterized by a general set $(g,\nu)$.
In the bosonic formulation, the model is still described by the
Lagrangian \eqref{jssksjsak}, where
the  couplings $(a_1,a_2)$  substitute the pair 
$(g,\nu)$. 
These two pairs of renormalization group invariants
are related as follows \cite{Bukhvostov:1980sn,Fateev:1996ea}:\footnote{Here, again, it is  assumed that we are dealing with
the  regularization of the
fermionic theory which preserves the $U(1)\otimes U(1)$ invariance.}
\bea\label{aikallaalsi}
\nu=\frac{1}{2}\ (a_1+a_2-2)\ ,\ \  \ \ \ \ \ \
\frac{g}{\pi}=\frac{a_2-a_1}{2 a_1 a_2}\ .
\eea
Due to the periodicity of the potential term  in $\varphi_i$, the space of states splits into
the orthogonal subspaces ${\cal H}_{\bf k}$ characterized by 
two ``quasimomenta'' ${\bf k}=(k_1,k_2)$,
\bea
\varphi_i\mapsto \varphi_i+\frac{4\pi}{\sqrt{a_i}}\ :\ \ \ \ 
|\,\Psi_{\bf k}\,\rangle\mapsto \re^{2\pi\ri k_i}\, |\,\Psi_{\bf k}\,\rangle\ .
\eea
As usual in the bosonization, the quasimomenta
are  related to the fermionic twists
\eqref{apssspps}:
\bea
k_\pm=\frac{1}{2}\ (k_1\pm k_2)\ .\label{apssspps2}
\eea
The neutral (w.r.t. $U(1)\otimes U(1)$) sector of the theory is described by   the Bose fields
with  periodic  boundary conditions:
\bea\label{bound}
\varphi_i(x^0,x^1+R)=\varphi_i(x^0,x^1)\ .
\eea
In the Euclidean version of \eqref{jssksjsak}, the periodic boundary 
corresponds to the geometry of infinite (or very
long in the``time'' direction $x^0$) flat cylinder
\bea\label{oasiosai}
D=\big\{{\bf x}=(x^0,x^1)\,|\, -\infty<x^0<\infty,\ 
x^1\equiv  x^1+R\big\}\ .
\eea
Then  the  ratio $E_{\bf k}/R$  would correspond to the specific (per unit length of
the cylinder) free energy with the scalar operator $\exp\big(\ri (k_1\varphi_1+k_2\varphi_2)\big)$
``flowing'' along the cylinder. The UV conformal dimension of this operator
is $\Delta=\frac{1}{4}\sum_{i=1}^2a_ik_i^2$. Therefore, we expect that at $R\to 0$
\bea\label{aosiaioas} 
E_{\bf k}\sim -\frac{\pi}{6R}\ c_{\bf k}\ ,\ \ \ \ \ c_{\bf k}=\sum_{i=1}^2\big(1-6 a_ik_i^2\big)\ .
\eea
The conformal perturbation theory  for $E_{\bf k}$ is constructed in
the usual way \cite{Zamolodchikov:1994uw} and yields an expansion
in the dimensionless variable $\lambda=2\pi\mu\, \big(\frac{R}{2\pi}\big)^{1-\nu}$,
\bea\label{aasioaisaasi}
E_{\bf k}=\frac{\pi}{R}\  
\sum_{n=0}^\infty e_{n}^{(\nu)}\ \lambda^{2n}\ .
\eea
Here the first coefficient  $e^{(\nu)}_0$  coincides with $-\frac{c_{\bf k}}{6}$, while 
the subsequent ones  are given by the perturbative integrals.
In particular $e_1^{(\nu)}=I(p_+)+I(p_-)$, where $p_\pm=\frac{1}{2}(a_1k_1\pm  a_2k_2)$ and
\bea
\label{sssai}
I(p)&=& \int_D\frac{\rd^2 x}{ R^2}
\frac{ 4^{-\nu}\, \pi\, \re^{-\frac{2\pi}{R} (\nu+2p) x^0}}{
\big(\sinh\big(\frac{\pi}{R}(x^0+\ri x^1)\big)\sinh\big(\frac{\pi}{R}(x^0-\ri x^1)\big)\big)^{1+\nu}}\nonumber
\\[.3cm] 
&=&
\frac{\Gamma(\frac{1}{2}+p +\nu) \Gamma(\frac{1}{2}-p)\Gamma(-\nu)}
{\Gamma(\frac{1}{2}-p-\nu) \Gamma(\frac{1}{2}+p) \Gamma(1+\nu)}\ .
\eea 

In the opposite  large-$R$ limit,
the vacuum energy is composed of an extensive part
which is proportional
to the spatial size of the system and does not depends on
the quasimomenta.
The specific bulk energy, ${\cal E}\equiv\lim_{R\to\infty}E_{\bf k}/R$, has  dimension $[\,mass\,]^2$, i.e.,
${\cal E}/M^2$ is a certain function
of  the  dimensionless couplings $(g,\nu)$.  This universal ratio, along
with another
dimensionless combinations  $\mu/M^{1-\nu}$,
are fundamental  characteristic of the theory, which allows one to
glue together the small- and large-$R$  asymptotic  expansions.
It is convenient to 
extract the extensive part from $E_{\bf k}$ and introduce
the scaling function
\bea\label{osaoasisoais}
{\mathfrak F}(r,{\bf k})=\frac{R}{\pi}\ ( E_{\bf k}-R\ {\cal E})\ .
\eea
Notice that
it is a  dimensionless function of the dimensionless variables  $r\equiv MR$ and ${\bf k}$ (and, of course,
the couplings),
satisfying the normalization condition
\bea\label{aosioaioasas}
\lim_{r\to+\infty}{\mathfrak F}(r,{\bf k})= 0\ .
\eea
Also, since  the   value of  $c_{\rm eff}\equiv-6\, {\mathfrak F}(r,{\bf k})$ at $r=0$
coincides with
the UV effective central charge  \eqref{aosiaioas},
this function can be interpreted as an effective central charge 
for the off-critical theory.

After this preparation let us turn to the case $\nu=0$.
Now, as it follows from  the relations \eqref{aikallaalsi}, the parameters of the bosonic
Lagrangian \eqref{jssksjsak}
obey the constraint
\bea 
a_1+a_2=2\ ,
\eea
which can be resolved as
\bea
a_1=1-\delta\ ,\ \ \ a_2=1+\delta\ .
\eea
We will assume that $0<a_1\leq 1\leq a_2$, i.e.,
$0\leq \delta<1$.
A formal substitution of $\nu=0$  in  \eqref{sssai} leads to a
divergent  expression. In order 
to regularize $I(p)$, we
cut a small
disk $|{\bf x}|<\epsilon$  in the  integration domain $D$.
As $\epsilon\to 0$, the regularized integral 
diverges logarithmically:
\bea\label{saisssaoo}
I^{(\epsilon)}(p)|_{\nu=0}=-2\, \log\big({\textstyle \frac{2\pi}{R}}\, \epsilon\big)
-\psi\big({\textstyle\frac{1}{2}}+p\big)-\psi\big({\textstyle\frac{1}{2}}-p\big)-2\gamma_{\rm E}+o(1)\ ,
\eea
where $\psi$ stands for the logarithmic derivative of the $\Gamma$-function and $\gamma_{\rm E}$ is the 
Euler constant.
In the case $\nu=0$, the general small-$R$ expansion is substituted by the
asymptotic series  of the form
\beq\label{aolissiao}
\frac{RE_{\bf k}}{\pi}\asymp -\frac{1}{3}+\frac{4p_1^2}{1-\delta}+\frac{4p_2^2}{1+\delta}-
(\mu R)^2\ \Big(e_1(0)-4\,
\log\big({\textstyle \frac{2\pi}{R}}\, \epsilon\,\re^{\gamma_E-\frac{1}{2}}\big)\Big)-
\sum_{n=2}^\infty e_n(\delta)\ (\mu R)^{2n}\ ,
\eeq
where explicitly
\bea\label{apspaspaspas}
e_1(0)=-2-\psi\big({\textstyle\frac{1}{2}}+p_1+p_2\big)-\psi\big({\textstyle\frac{1}{2}}-p_1-p_2\big)-
\psi\big({\textstyle\frac{1}{2}}+p_1-p_2\big)-\psi\big({\textstyle\frac{1}{2}}-p_1+p_2\big)
\eea
and
\bea
\label{kasoisau}
p_1=\frac{1}{2}\ (1-\delta)\ k_1\ ,\ \ \  \ \ 
p_2=\frac{1}{2}\ (1+\delta)\ k_2\ .
\eea

In ref.\,\cite{Fateev:1996ea} Fateev presented strong arguments
supporting the integrability of the  BL model with $\nu=0$ and 
found  an exact $\mu-M$ relation,
\bea\label{jshsyusy}
\mu={\textstyle
  \frac{M}{2\pi}}\ \cos\big({\textstyle\frac{\pi\delta}{2}}\big)\ . 
\eea
Using  his results  it is straightforward to obtain 
(see Sec.\,\ref{secnew5} bellow) the
following expression for the bulk specific energy
\bea\label{oaisaiasoso}
{\cal E}=\pi\mu^2\ \Big(\,4\,
\log\big( \pi\mu\epsilon\,\re^{\gamma_{\rm E}-\frac{1}{2}}\big)+
\psi\big({\textstyle\frac{1+\delta}{2}}\big)+
\psi\big({\textstyle\frac{1-\delta}{2}}\big)-2\, \psi\big({\textstyle\frac{1}{2}}\big)\Big)\ .
\eea
One can see now that  ${\mathfrak F}$, defined by eq.\,\eqref{osaoasisoais},
does not contain any UV divergences, i.e.,
it is an universal  scaling function of the dimensionless variable  $r=MR$.
Its  small-$R$
expansion can be written in the form
\bea\label{apspassp}
{\mathfrak F}(r,{\bf k})\asymp -\frac{1}{3}+2k_+^2+2k_-^2-4\delta\,k_+k_--
16\,\rho^2\, \log(\rho)-
\sum_{n=1}^\infty e_n(\delta)\ (2\rho)^{2n}\ ,
\eea
where $k_\pm=\frac{1}{2}(k_1\pm k_2)$,
$\rho={\textstyle \frac{r}{4\pi}}\ \cos\big({\textstyle\frac{\pi\delta}{2}}\big)$ and
\bea\label{aosiosaaso}
e_1(\delta)=e_1(0)+\psi\big({\textstyle\frac{1+\delta}{2}}\big)+
\psi\big({\textstyle\frac{1-\delta}{2}}\big)-2\, \psi\big({\textstyle\frac{1}{2}}\big)\ .
\eea

A few comments are  in order here.
As it was already mentioned in the introduction,
the logarithmic divergence of ${\cal E}$ is
well expected  in the context of  application of the BL model to the problem of 
instanton summation in the $O(3)$ sigma model. 
The integration over the instanton moduli space leads
to the divergent contribution of the small-size instantons \cite{Luscher:1981tq}. 
So that $\epsilon$ can be interpreted as a cut-off parameter
which allows one to exclude the divergent contribution of the small-instantons.
Another comment  concerns to the symbol $\asymp$,  which is used in eqs.\,\eqref{aolissiao}
and \eqref{apspassp}
 to emphasize the  asymptotic nature of these 
power series expansions.
To see that they  
have zero radius of convergence,
it is sufficient to consider the case $\delta=0$.
Returning to the fermionic description, the   model \eqref{Lagr1} with $g=0$ constitutes a
pair of non-interaction Dirac fermions, so that 
there exists a  closed analytic expression for the scaling function 
${\mathfrak F}_{0}={\mathfrak F}|_{\delta=0}$. Namely,
$
{\mathfrak F}_0(r,{\bf k})=
{\mathfrak f}(r,k_+)+ {\mathfrak f}(r,k_-)$,
where $\pi {\mathfrak f}/R^2 $ coincides  with
the specific free energy
 of the free  Dirac fermion at
the temperature $1/R$ and (imaginary)  chemical potential $2\pi\ri k/R$, i.e.,
\bea\label{apoasoasp}
{\mathfrak f}\big(r,k)=
-\frac{r}{2\pi^2}\ \int_{-\infty}^\infty\rd\theta\,\cosh(\theta)
\ \log\Big[\Big(1+\re^{2\pi\ri k}\re^{-r\cosh(\theta)}\Big)
\Big(1+\re^{-2\pi\ri k}\re^{-r\cosh(\theta)}\Big)\,\Big]\ .
\eea
It is now straightforward to see that 
the power series  \eqref{apspassp} for $\delta=0$  is indeed an  asymptotic expansion and  
\bea
e_n(0)&=&-2\,\delta_{n,1}+
\frac{(-1)^n\,n}{4^{n-1} (n!)^2}\ 
\  \Big(\, \psi^{(2n-2)}\big({\textstyle \frac{1}{2}}+p_1+p_2\big)+
                    \psi^{(2n-2)}\big({\textstyle \frac{1}{2}}-p_1-p_2\big)\nonumber
\\
&+&
                    \psi^{(2n-2)}\big({\textstyle \frac{1}{2}}+p_1-p_2\big)+
                    \psi^{(2n-2)}\big({\textstyle \frac{1}{2}}-p_1+p_2\big)\,\Big)\ ,
\eea
where  the superscript  stands for  derivative  of $(2n-2)$-order w.r.t. the argument.

\begin{figure}
\centering
\includegraphics[width=9.5cm]{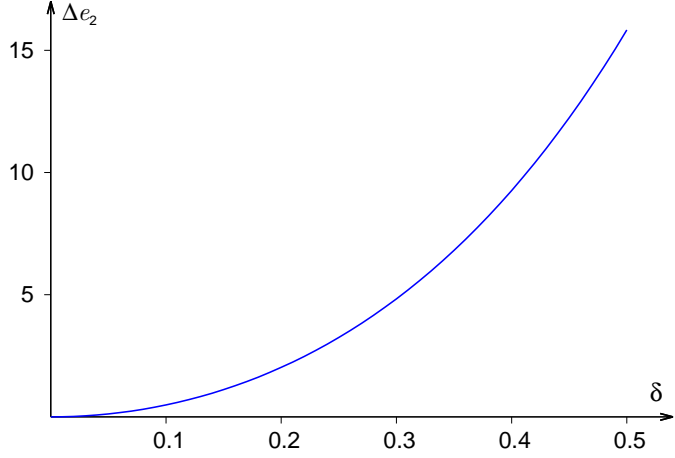}
\caption{
The difference 
$\Delta e_2\equiv e_2(\delta)-e_2(0)$, defined by \eqref{aasopaps}, as
a function of $\delta$ for $k_1=k_2=0$. Note that in this case 
$\Delta e_2=48.21714061416\ldots\times \delta^2
+
60.427986409885\ldots\times\, \delta^4+O(\delta^6)$ as $\delta\to 0$.
} 
\label{fig6bb}
\end{figure}

For nonzero $\delta$ the asymptotic coefficients $e_n(\delta)$ with $n\geq 2$ can be
expressed in terms of the multiple integrals. Unfortunately such representation
can not be used for any practical purposes. The only exclusion is  $e_2(\delta)$, whose
integral representation can be simplified 
dramatically. For future references we describe here major steps in 
this calculation.
First of all, using the complex coordinate $z=\exp(2\pi (x^0+\ri x^1)/R)$,
the asymptotic coefficient $e_2(\delta)$ can be represented as a
6-fold integral, 
\bea\label{aosiosoaaa}
e_2(\delta)&=& e_2(0)+
2\ \int\prod_{i=1}^3\frac{\rd^2 z_i}{2\pi}\
|z_1|^{-1+2p_1+2p_2}\
|z_2|^{-1-2 p_1+2p_2}\
|z_3|^{-1-2p_1-2p_2}\
\nonumber\\
&\times&
\bigg(\,\bigg|\frac{(z_1-z_2)(1-z_3)}{(z_3-z_2)(1-z_1)}\bigg|^{2 \delta}-1\bigg)\
\big|(1-z_2)(z_1-z_3)\big|^{-2}\ .
\eea
Now let us substitute the integration variable $z_2$
with $\zeta=\frac{(1-z_1)(z_2-z_3)}{(1-z_2)(z_1-z_3)}$, and 
integrate over $z_1$  by means of  the  identity
\bea\label{opasssaaspsp}
\int\frac{\rd^2 z}{\pi}\, |z|^{-1+2p_1+2p_2}\, |1-z|^{-1+2 p_1-2p_2}\, 
|z-w|^{-1-2p_1+2p_2}=
|w|^{-1+2p_2}\ |1-w|\  \tau_{p_1 p_2}\big({\textstyle\frac{1}{1-w}}\big)\ .
\eea
Here 
\bea\label{tau12}
\tau_{p_1p_2}(\zeta)
&=&
\frac{\Omega(-p_1,p_2) }{2p_1}\
|\zeta |^{1-2p_1}|1-\zeta|^{1+2 p_2}
\big|{}_2F_1\big({\textstyle \frac{1}{2}}- p_1+p_2,
{\textstyle \frac{1}{2}}- p_1+p_2, 1-2 p_1;\zeta\big)\big|^2\nonumber\\
&-&\frac{\Omega(p_1,p_2) }{2p_1}\
|\zeta|^{1+2p_1}|1-\zeta|^{1-2 p_2}
\big|{}_2F_1\big({\textstyle \frac{1}{2}}+ p_1-p_2,
{\textstyle \frac{1}{2}}+ p_1-p_2, 1+2 p_1;\zeta\big)\big|^2 ,
\eea
${}_2F_1$ stands for the conventional hypergeometric function,
and
\bea
\Omega(p_1,p_2)=\frac{\Gamma(\frac{1}{2}+p_1-p_2)\Gamma(\frac{1}{2}+p_1+p_2)}
                {\Gamma(\frac{1}{2}-p_1-p_2)\Gamma(\frac{1}{2}-p_1+p_2)}\ 
\frac{\Gamma(1-2p_1)}{\Gamma(1+2p_1)}\ .
\eea
Finally, the integral over $z_3$ can be performed using a remarkable
relation 
\bea\label{ahssaajh}
\big(\tau_{p_1 p_2}(\zeta)\big)^2=
|\zeta|^2\ \int
\frac{\rd^2 z}{\pi|z|^2}\
\Big|\frac{1-\zeta z }{z (z-\zeta)}\Big|^{2p_1}\
\frac{\tau_{p_1 p_2}\big(X(z)\big)}{|X(z)|}\ ,
\eea
where $X(z)=\frac{(\zeta-z)(1-\zeta z)}{\zeta(1-z)^2}$.
As a final  result one obtains  the following integral representation
\bea
\label{aasopaps}
e_2(\delta)=e_2(0)+\frac{1}{4\pi}\ \int\frac{\rd^2\zeta}{|\zeta|^{2}|1-\zeta|^{2}}\ \Big(
|\zeta|^{-2\delta}|1-\zeta|^{2\delta}-1\Big)\ \tau^2_{p_1 p_2}(\zeta)\ .
\eea
This formula allows one  to achieve a  reliable accuracy
in the  numerical calculation of $e_2(\delta)$.
For  illustration,  we present in Fig.\ref{fig6bb} the numerical  results 
for  $k_1=k_2=0$.
Notice that
in this case
the corresponding function
$\tau_{00}(\zeta)$ in  \eqref{aasopaps} 
can be expressed in terms of 
the complete elliptic
integral of the first order $K(\zeta)=
\frac{\pi}{2}\ {}_2F_{1}(\frac{1}{2},\frac{1}{2},1;\zeta)$:
\bea
\tau_{00}(\zeta)=\frac{8}{\pi}\ |\zeta(1-\zeta)|\ 
\Re e\big(K^*( \zeta)K(1-\zeta)\big)\ .
\eea

Note that $\tau_{p_1 p_2}(\zeta)$ given in \eqref{tau12} is a
particular case of a more general function $\tau_{p_1 p_2 p_3}(\zeta)$
defined by \eqref{oaissisaoi}, namely $\tau_{p_1 p_2}(\zeta)\equiv
\tau_{p_1 p_2 p_3}(\zeta)|_{p_3=0}$. This function defines a real solution
\eqref{liouville1} of the Liouville equation \eqref{liouville2},
satisfying the asymptotic conditions \eqref{paspapsoasa}.

\section{\label{sec3}Weak coupling expansion}

We now consider a  weak  coupling  expansion of the scaling function ${\mathfrak F}$.
Since $g=\frac{\pi\delta}{1-\delta^2}=
\pi\delta+O(\delta^3)$,
no needs to distinguish  $\frac{g}{\pi}$ and $\delta$
within  the first  two  perturbative orders. It is convenient to
define
the perturbative coefficients   through the relation:
\bea\label{apospapaso}
{\mathfrak F}={\mathfrak F}_{0}
+ {\mathfrak F}_{1}\, \delta+
{\mathfrak F}_{2}\, \delta^2+O(\delta^3)\ .
\eea
Here   ${\mathfrak F}_0(r,{\bf k})={\mathfrak f}(r,k_+)+ {\mathfrak f}(r,k_-)$ with
${\mathfrak f}$ given  by \eqref{apoasoasp} (recall  that $k_\pm=\frac{1}{2}\,(k_1\pm k_2)$).
The results  obtained in the previous section allows one to predict the
leading small-$R$ behavior of ${\mathfrak F}_{i}$. 
Generally speaking the  coefficients in the power series \eqref{apspassp}
admit the Taylor expansion
$
e_n(\delta)=e_n(0)+ \big(\frac{\partial e_n(0)}{\partial \delta}\big)_{p}\ \delta+
\frac{1}{2}\ \big(\frac{\partial^2 e_n(0)}{\partial \delta}\big)_{p}\ 
\delta^2+O(\delta^3)$.
In particular, as it follows from eq.\eqref{aosiosaaso},
\bea
{\textstyle
\big(\frac{\partial e_1(0)}{\partial \delta}\big)_{p}=0\ ,\ \ \ \ 
\big(\frac{\partial e_1(0)}{\partial \delta}\big)_{p}=\frac{1}{2}\ 
\psi''''\big({\textstyle\frac{1}{2}}\big)\ .}
\eea
Also, using the  original integral representation \eqref{aosiosoaaa} for
$e_2(\delta)$, one can show that
\bea
\big({\textstyle \frac{\partial e_2(0)}{\partial \delta}}\big)_{p}=
{\textstyle\frac{1}{4}}\, \big(\, \psi'\big({\textstyle \frac{1}{2}}+p_1+p_2\big)-
                    \psi'\big({\textstyle \frac{1}{2}}-p_1-p_2\big)\big)\,\big(
                    \psi'\big({\textstyle \frac{1}{2}}+p_1-p_2\big)-
                    \psi'\big({\textstyle \frac{1}{2}}-p_1+p_2\big)\,\big)\, .
\eea
In the case $p_1=p_2=0$ the weak coupling expansion includes only  even powers of $\delta$ (see Fig.\,\ref{fig6bb}) and
\bea
{\textstyle 
\big(\frac{\partial^2 e_2(0)}{\partial \delta^2}\big)_{p_1=p_2=0}=-{\textstyle\frac{1}{8}}\ 
\psi''''\big({\textstyle\frac{1}{2}}\big)\ .}
\eea
All of  these   can be used to study the short distance expansion of ${\mathfrak F}_{i}$
in eq.\eqref{apospapaso}.\footnote{ 
Recall that  
the relations \eqref{kasoisau} between $k_i$ and $p_i$ involve  the perturbative coupling. 
This should be  taken into account since  it is assumed that the expansion \eqref{apospapaso} is
performed for fixed values of $k_i$ rather then $p_i$.}
In particular, it is possible to show that
\bea\label{papospasoa}
{\mathfrak F}_1(r,{\bf k})=- 4\,{\mathfrak q}(r,k_+)\, {\mathfrak  q}(r,k_-)\ ,
\eea
where 
\beq\label{paospspasopsa}
{\mathfrak  q}(r,k)= k+
\big(\psi'\big({\textstyle\frac{1}{2}} +k\big)-\psi'\big({\textstyle\frac{1}{2}} -k\big)\big)
\Big({\frac{r}{4\pi}}\Big)^2-{\frac{1}{2}}\,
\big(\psi'''\big({\textstyle\frac{1}{2}} +k\big)-
\psi'''\big({\textstyle\frac{1}{2}} -k\big)\big)\, \Big({\frac{r}{4\pi}}\Big)^4
+O(r^6) 
\eeq
and also
\bea
{
{\mathfrak F}_{2}= \frac{r^2}{4}\ \log\Big(\frac{r}{4\pi}\Big)
+A\  \Big({\frac{r}{4\pi}}\Big)^2
-
B\ \Big({\frac{r}{4\pi}}\Big)^4+O(r^6)\ ,}
\eea
where
\bea
A&=&-\psi''\big({\textstyle\frac{1}{2}}\big)-
\pi^2\ \Big(\psi\big({\textstyle\frac{1}{2}} +k_+\big)+\psi\big({\textstyle\frac{1}{2}} +k_-\big)+
\psi\big({\textstyle\frac{1}{2}} -k_+\big)+\psi\big({\textstyle\frac{1}{2}} -k_-\big)\,\Big)\\
&+&
2 k_+^2\,
\Big(\psi''\big({\textstyle\frac{1}{2}} +k_-\big)+\psi''\big({\textstyle\frac{1}{2}} -k_-\big)\,\Big)+
2k^2_-\,
\Big(\psi''\big({\textstyle\frac{1}{2}} +k_+\big)+\psi''\big({\textstyle\frac{1}{2}} -k_+\big)\,\Big)\ .
\nonumber
\eea
In the case  $k_1=k_2=0$, 
\bea\label{asaposoosap}
A|_{k_1=k_2=0}= -\psi''\big({\textstyle\frac{1}{2}}\big)-
4\pi^2\, \psi\big({\textstyle\frac{1}{2}}\big)\ ,\ \ \  \ 
B|_{k_1=k_2=0}=-\psi''''\big({\textstyle\frac{1}{2}}\big)-4\pi^2\, \psi''\big({\textstyle\frac{1}{2}}\big)\ .
\eea

\begin{figure}
\centering
\psfrag{a}{$-$}
\psfrag{b}{$+$}
\includegraphics[width=4.5cm]{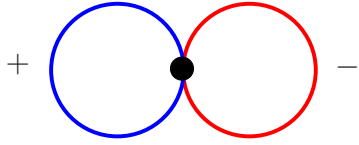}
\caption{A diagrammatic representation of ${\mathfrak F}_1$ in  eq.\,\eqref{apospapaso}.
The signs $\pm$ label
the  fermion ``colors'' $\psi_\pm$ propagating  along
the  loops (see   Lagrangian \eqref{Lagr2}).
}
\label{fig2a}
\end{figure}

For finite values of $r$ 
the perturbative coefficients ${\mathfrak F}_i$ can be calculated within 
the renormalized perturbation theory based on  Lagrangian \eqref{Lagr2}.
Let 
${\boldsymbol S}_\sigma({\bf x})\equiv \langle\, \psi_\sigma ({\bf x})\otimes {\bar \psi}_\sigma({\bf 0})\,\rangle
$ (${\bf x}=(x^0,x^1),\,\sigma=\pm)$
be the fermionic Matsubara  propagator with the  temperature $1/R$ and
chemical potential $2\pi\ri k_\sigma/R$.
It  can be expressed  in terms of the 
modified Bessel function of the second kind
$K_s(z)=\frac{1}{2}\ \int_{-\infty}^\infty\rd\theta\ \re^{s\theta-z\cosh(\theta)}$,
\bea
{\boldsymbol S}_\sigma({\bf x})=\big(M-\gamma^a\partial_a\big)\ G_\sigma({\bf x})\ ,
\eea
where $\gamma^a$ are Euclidean $\gamma$-matrices, $\{\gamma^a,\gamma^b\}=2\,\delta^{ab}$,
and
\bea
G_\sigma({\bf x})=\frac{1}{2\pi}\ \sum_{n=-\infty}^\infty(-1)^{n}\ 
\re^{2\pi\ri n k_\sigma }\  K_0\big(|w-\ri\,n r|\big)\ \ \  \ \ \  \ {\rm with}\ \ \ \ \ 
w=M\, (x^0+\ri\,x^1)\ .
\eea
At the first perturbative order one has (see Fig.\,\ref{fig2a})
\bea
{\mathfrak F}_1=
R^2\ 
{\rm Tr}\big({\boldsymbol S}_+(0)\gamma_a\big)\,{\rm Tr}\big({\boldsymbol S}_-(0)\gamma^a\big)=
4R^2\ 
\langle\, \Psi^\dagger_+\, \Psi_+(0)\,\rangle\, 
\langle\, {\bar \Psi}^\dagger_-\, {\bar \Psi}_-(0)\,\rangle\ .
\eea
Here
$\Psi_\sigma$ and ${\bar \Psi}_\sigma$ stand  for  the components of the
Dirac  bispinors $\psi_\sigma$  with the Lorentz spin $+\frac{1}{2}$ and $-\frac{1}{2}$, respectively.
In zero-temperature limit 
the Lorentz invariance is  restored and hence 
$
\langle\, {\bar \Psi}^\dagger_\sigma\, {\bar \Psi}_\sigma(0)\,\rangle=
-\langle\, \Psi^\dagger_\sigma\, \Psi_\sigma(0)\,\rangle\to 0$.
Introducing function ${\mathfrak q}$ through the relation
\bea\label{paopsopas}
\langle\, {\bar \Psi}^\dagger_\sigma\, {\bar \Psi}_\sigma(0)\,\rangle=-
\langle\, \Psi^\dagger_\sigma\, \Psi_\sigma(0)\,\rangle=
\frac{1}{R}\ {\mathfrak q}(r,k_\sigma)\ ,
\eea
one observes that  ${\mathfrak F}_1$
takes the form
\eqref{papospasoa}.
It is also easy to see that 
\bea\label{laaaksussa}
{\mathfrak q}=\frac{1}{4}\ \frac{\partial {\mathfrak f}}{\partial k}\ ,
\eea
where ${\mathfrak f}={\mathfrak f}(r,k)$ is given by \eqref{apoasoasp}.
This is in a complete  agreement with the short distance prediction \eqref{paospspasopsa}.

\begin{figure}
\centering
\psfrag{a}{$+$}
\psfrag{b}{$-$}
\psfrag{c}{$-\sigma$}
\psfrag{d}{$\sigma$}
\psfrag{G}{${\rm (I)}$}
\psfrag{H}{${\rm (II)}$}
\psfrag{W}{${\rm (III)}$}
\includegraphics[width=16cm]{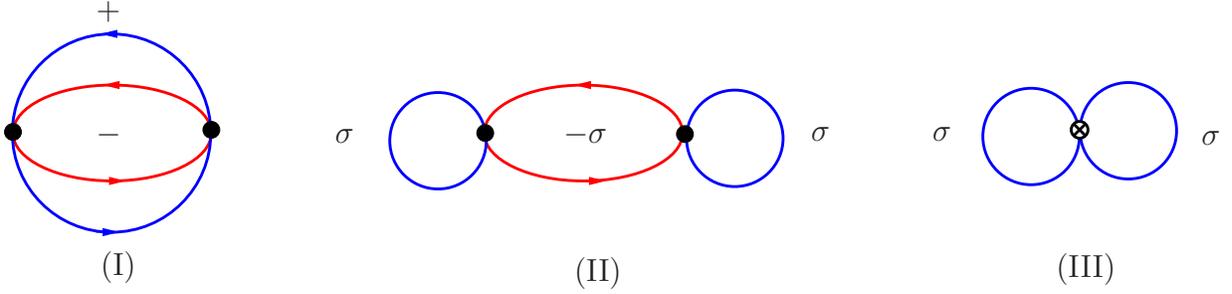}
\caption{The diagrams contributing to
the second perturbative order. 
The contribution of the counterterm  $\propto g_1$  in \eqref{Lagr2} is
visualized by the type III diagrams   (as $\nu=0$,
there is no mass renormalization, i.e.   $\delta M=0$).
}
\label{fig3a}
\end{figure}

The second-order diagrams are depicted in Fig.\,\ref{fig3a}.
The  type I diagram gives the contribution
\bea\label{appsopsaoyss}
{\mathfrak F}_2^{\rm (I)}=-\frac{\pi}{2}\ r^2\, \int_{D_\epsilon}\rd^2x\ 
{\rm Tr}\big({\boldsymbol S}_+(-{\bf x})\,\gamma_a\,
{\boldsymbol S}_+({\bf x})\, \gamma_b\big)\
{\rm Tr}\big({\boldsymbol S}_-(-{\bf x})\,\gamma^a\,
{\boldsymbol S}_-({\bf x})\, \gamma^b\big)\ .
\eea
Because of the UV  divergence  at ${\bf x}=0$,
the  integration domain $D_\epsilon$ here
 is chosen to be the cylinder \eqref{oasiosai}
without   an  infinitesimal hole   $|{\bf x}|<\epsilon$.
One can show that, as  $\epsilon$  tends to zero,
\bea
{\mathfrak F}^{\rm (I)}_2=-\Big(\frac{R}{2\pi\epsilon}\Big)^2+
\sum_{\sigma=\pm}\Big( {\mathfrak t}(r,k_\sigma)-
\frac{r}{2\pi}\, \log\big({\textstyle\frac{M\epsilon}{2}}\, \re^{\gamma_E-
\frac{1}{2}}\big)
\Big)^2+{\rm finite}\ ,
\eea
where
\bea
{\mathfrak t}=-\pi\ 
\frac{\partial {\mathfrak f}}{\partial r}\ .
\eea
In fact, since $E_{\bf k}=R\,{\cal E}+\frac{\pi}{R}\, {\mathfrak F}$, 
the  quadratic divergence  $\propto 1/\epsilon^2$ 
should be relocated to
the  specific bulk energy.
Generally speaking, the specific bulk energy has a form
\bea\label{paopsaoaops}
{\cal E}= 
w(g)\, \Lambda^2+\frac{M^2}{\pi}\, \cos^2\big({\textstyle\frac{\pi\delta}{2}}\big)\ 
\log\big(M/\Lambda\big)+o(1)\ ,
\eea
where $\Lambda\gg M$ is some lattice energy scale and
$w$ is some (nonuniversal) function of  the coupling  $g$. 
Notice that, in writing eq.\,\eqref{oaisaiasoso}, the quadratic divergence
was omitted (as usual in QFT).

The type II diagrams from  Fig.\,\ref{fig3a} leads to  the UV finite
integral over the whole cylinder  $D$:
\bea
{\mathfrak F}^{\rm (II)}_2=\frac{\pi}{2}\ r^2\, \int_{D}\rd^2x\ 
\sum_{\sigma=\pm} {\rm Tr}\big({\boldsymbol S}_{\sigma}(0)\,\gamma_a\big)\,
{\rm Tr}\big({\boldsymbol S}_{-\sigma}(-{\bf x})\,\gamma^a\,
{\boldsymbol S}_{-\sigma}({\bf x})\, \gamma^b\big)\, 
{\rm Tr}\big({\boldsymbol S}_{\sigma}(0)\,\gamma_b\big)
 \ .
\eea

\begin{figure}
\centering
\includegraphics[width=8cm]{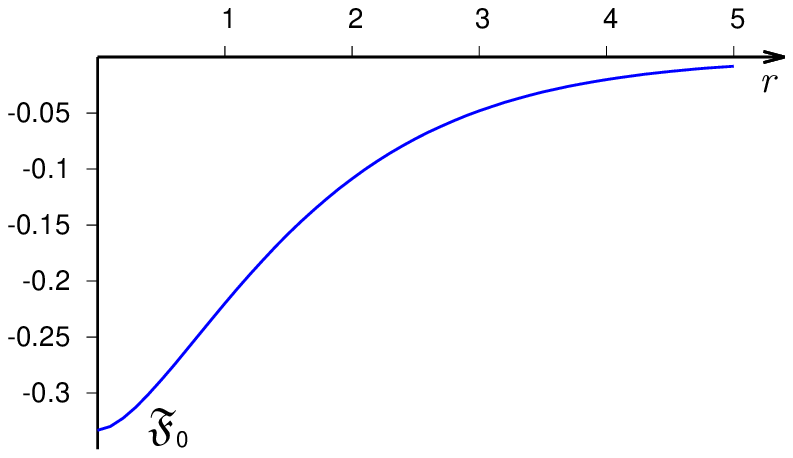}
\includegraphics[width=8cm]{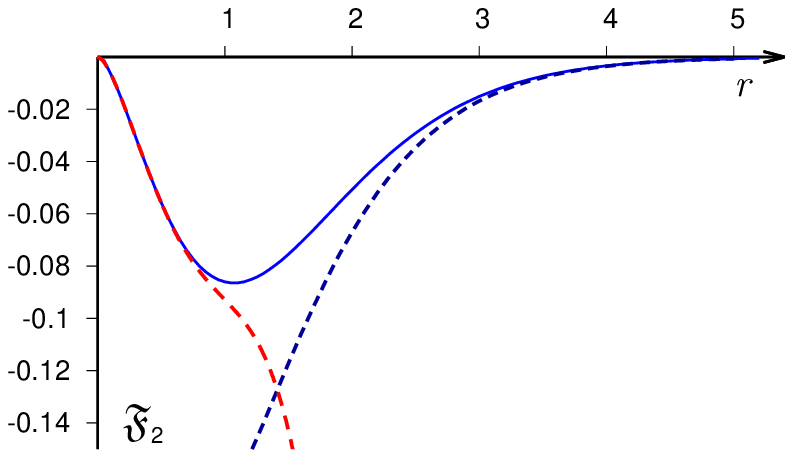}
\caption{The perturbative coefficient ${\mathfrak F}_{i}$ \eqref{apospapaso} 
for $k_1=k_2=0$ $({\mathfrak F}_{1}=0$ in this case).
The left panel shows ${\mathfrak F}_{0}=
-\frac{2r}{\pi^2}\ \int_{-\infty}^\infty\rd\theta\,\cosh(\theta)
\, \log\big(1+\re^{-r\cosh(\theta)}\big)$.
At the right panel ${\mathfrak F}_{2}$ is compared against
its
large-$r$ asymptotic ${\mathfrak F}_{2}=-\frac{3}{2}\,r^2\,K_0(2r)+o(\re^{-2r})$ (blue dashed line)
and the small-$r$ asymptotic ${\mathfrak F}_{2}= \frac{r^2}{4} \, 
\log\big(\frac{r}{4\pi}\big)+A\, \big(\frac{r}{4\pi}\big)^2-B\, \big(\frac{r}{4\pi}\big)^4+O(r^6)$
(red dashed line). The numerical coefficients $A$ and $B$ are given by eq.\eqref{asaposoosap}.
}
\label{fig4}
\end{figure}

Finally, the 
counterterm $\propto g_1$ in  \eqref{Lagr2} contributes through the type III diagrams,
schematically visualized in
Fig.\,\ref{fig3a}. This can be written in the form
$ \frac{2g_1}{\pi}\ {\mathfrak F}^{\rm (III)}_2 $ with
\bea\label{aapasosaosa}
{\mathfrak F}^{\rm (III)}_2&=&
\frac{1}{4}\ R^2\ \sum_{\sigma=\pm}\Big(
{\rm Tr}\big({\boldsymbol S}_\sigma (0)\gamma^a\big)\,
{\rm Tr}\big({\boldsymbol S}_\sigma(0)\gamma_a\big)
-{\rm Tr}\big({\boldsymbol S}_\sigma(0)\gamma^a{\boldsymbol S}_\sigma(0)\gamma_a\big)\Big)\nonumber\\
&=&R^2\sum_{\sigma=\pm }\big(\langle\, \Psi^\dagger_\sigma\, \Psi_\sigma(0)\,\rangle
\langle\, {\bar \Psi}^\dagger_\sigma\, {\bar \Psi}_\sigma(0)\,\rangle- 
\langle\, {\bar \Psi}_\sigma\, \Psi^\dagger_\sigma(0)\,\rangle^2\big)\ .
\eea
Contrary to  the one point functions \eqref{paopsopas}, the condensate
$\langle\, {\bar \Psi}_\sigma\, \Psi^\dagger_\sigma(0)\,\rangle$ diverges logarithmically:
\bea\label{aaoisosaoa}
\langle\, {\bar \Psi}_\sigma\, \Psi^\dagger_\sigma(0)\,\rangle=
\langle\, { \Psi}_\sigma\, {\bar \Psi}^\dagger_\sigma(0)\,\rangle=
\frac{1}{R}\  {\mathfrak t}(r,k_\sigma)
-
\frac{M}{2\pi}\ \log\Big(\frac{M\epsilon}{2} \, \re^{\gamma_E-\frac{1}{2}+C}
\Big)\ ,
\eea
where $C$ is some constant.
Since 
\bea\label{spsopsop}
 {\mathfrak F}^{\rm (I)}_2+
{\mathfrak F}^{\rm (III)}_2+\Big(\frac{R}{2\pi\epsilon}\Big)^2=-\frac{r^2 }{\pi^2 }\ C\
\log(M\epsilon)+{\rm finite}\ ,
\eea
the UV  divergence $\propto \log^2(\epsilon)$ 
is  canceled  from   the sum of types I and  III  diagrams  if we choose
$g_1=\frac{g^2}{2\pi}+O(g^3)$.
As well as the quadratic  divergence,  the remaining logarithmic 
divergence    should be  relocated 
to the specific bulk energy. Expanding $\cos^2\big({\textstyle\frac{\pi\delta}{2}}\big)$
in \eqref{paopsaoaops} one can find  the  value of the constant $C$: 
\bea\label{aisiaosai}
C=\frac{\pi^2}{4}\ .
\eea
This way the second order correction takes  the form
\bea\label{apaspospa}
{\mathfrak F}_{2}=
\frac{r^2}{4\pi^2}\ {\cal C}_2+ 
\lim_{\epsilon\to 0}\bigg[\, \sum_{\alpha={\rm I,II,III}}{\mathfrak F}^{(\alpha)}_2+
\Big(\frac{R}{2\pi\epsilon}\Big)^2+
\frac{r^2 }{4 }\ \log\Big(\frac{M\epsilon}{2} \, \re^{\gamma_{\rm E}-\frac{1}{2}}\Big)
\,\bigg]\ ,
\eea
where the  finite constant should be adjusted to satisfy the normalization condition \eqref{aosioaioasas}.
It reads explicitly as
\bea
{\cal C}_2= \frac{\pi^4}{8}-\frac{1}{2}-\frac{1}{4}\ \psi''\big({\textstyle\frac{1}{2}}\big)\ .
\eea
Further calculations show that 
\bea\label{sauiuias}
{\mathfrak F}_{2}(r,{\bf k})&=&
-\frac{1}{2}\,\ \big(1+c(2k_1)+ c(2k_2)\big)\ \, r^2\,K_0(2 r)
\\
&-&\big(1-c(2k_1) c(2k_2)\big)\ r\,
\int_{-\infty}^\infty\frac{\rd \nu}{\pi}\ \frac{\nu^2\, K_{\ri\nu}(r)K_{1+\ri\nu}(r)}
{\sinh^2(\frac{\pi\nu}{2})}
+o\big(\re^{-2r}\big)\ .\nonumber
\eea
Here  the shortcut notation $c(k)=\cos(\pi k)$ is used and
$K_{s}(z)$ denotes   the modified Bessel function of the second kind:
\bea
K_s(z)=\frac{1}{2}\ \int_{-\infty}^\infty\rd\theta\ \re^{s\theta-z\cosh(\theta)}\ .
\eea
Also in eq.\,\eqref{sauiuias} and bellow, 
the symbol  $o\big(\re^{-2r}\big)$ denotes   a  remaining term  that  decays
faster than $r^{-N}\ \re^{-2r}$ for any positive $N$ as $r\to+\infty$. 
Notice that the  normalization condition
${\mathfrak F}_{2}=o\big(\re^{-r}\big)$
implies an absence of the finite renormalization  of the  fermion mass. 
It can be used for fixing the constant $C$ in \eqref{aaoisosaoa} and
hence
avoid any reference to
the exact relation \eqref{paopsaoaops}.
For $k_1=k_2=0$, the result of  perturbative   calculation
is presented  in Fig.\,\ref{fig4}.

Using eqs.\,\eqref{papospasoa} and \eqref{laaaksussa} it is 
easy to show that
\bea\label{oaaaaasisao}
{\mathfrak F}_{1}(r,{\bf k})
=\frac{2}{\pi^2}\ \big(c(2k_1)-c(2k_2)\big)
\  r^2\,K_1^2(r)+o\big(\re^{-2r}\big)\ .
\eea
Thus, at least at the first-two perturbative orders,  the leading large-$r$  behavior
of the scaling function ${\mathfrak F}$  is defined by ${\mathfrak F}_{0}$ only
and therefore
\bea\label{aosioasiasoas}
{\mathfrak F}(r,{\bf k})=-\frac{4}{\pi^2}\ c(k_1)\, c(k_2)\ r\,K_1(r)+o(\re^{-r})\ .
\eea
This  can be understood as follows.  The leading large-$R$ behavior comes from the
virtual fermions trajectories winding once around the Matsubara circle. Such trajectories
should be counted with the phase factor $\re^{\ri\pi (\sigma_1k_1+\sigma_2 k_2)}$
and, therefore,
the summation  over  four possible sign combinations 
with $\sigma_{1,2}=\pm 1$
gives rise 
eq.\eqref{aosioasiasoas}.
Thus we may expect that 
the asymptotic formula \eqref{aosioasiasoas}  holds true as the mass
of the first  bound state  $M_1=2M\,\cos(\frac{\pi\delta}{2})$  is greater than $M$, i.e., for  $\delta\in[0, \frac{2}{3})$.

Before concluding this section let us make a few remarks about the
(non-integrable) case with a non-zero value of $\nu$. Instead of adjusting the 
counterterm coupling $g_1$, 
the logarithmic divergences can be absorbed by the mass counterterm with 
$\delta M=\nu\ \log(M/\Lambda_{\rm UV})$, where 
$\nu=\frac{g^2}{\pi^2}-\frac{2 g_1}{\pi} $ and 
$\Lambda_{\rm UV}=\frac{2}{\epsilon}\ \exp(\frac{1}{2}-\gamma_{\rm E}-\frac{\pi^2}{4})$.
(This is an
infinitesimal  version of eq.\,\eqref{aoisasosa}  where  $M_0=M+\delta M$.)
As it was mentioned in the introduction,
the exponent $\nu$ and 
the four-fermion coupling $g$  can be thought of  as independent parameters for  the  family of BL models.
Using eq.\,\eqref{aapasosaosa}, it is easy to see that
\bea
\frac{\partial {\mathfrak F}}{\partial\nu}\Big|_{\nu=0}=
 \sum_{\sigma=\pm}\big( {\mathfrak q}^2(r,k_\sigma)+
{\mathfrak t}^2(r,k_\sigma)\big)+O(g)\ .
\eea
Finally we  note that for $\nu\not=0$, (an universal part of) the specific bulk energy
has a valid 
Laurent expansion
of the form
\bea
{\cal E}(g,\nu)=M^2\, \Big(\,h_{-1}(g)\, \nu^{-1}+\sum_{n=0}^\infty h_{n}(g)\ \nu^n\, \Big)\ ,
\eea
where  $h_{n}(g)$ admit  power  series  expansions in $g^2$.

\section{\label{secnew5}Exact formula for ${\mathfrak F}(r,{\bf k})$}

The BL model with  non-vanishing  $\nu=\frac{1}{2}\, (a_1+a_2-2)$ can be thought
as a sort  of analytical regularization of the model with $\nu=0$ --
the integrals  appearing in the conformal perturbation theory  converge for negative
values of $\nu$, but  become singular at $\nu\to 0^-$.
A brief inspection of eq.\,\eqref{sssai}
shows that a simple pole  $\frac{1}{\nu}$ replaces  the logarithmic divergence
$2\, \log(\epsilon \mu)+const$ in  \eqref{saisssaoo} which occurs  when the integral is  regularized by
excluding 
a  neighborhood  of the singular point  from the integration domain.
The BL with non-vanishing $\nu$ is a well defined  QFT and it is interesting in itself
in a context of applications in  condensed matter physics \cite{Lesage:1997jq}. 
However, as it was already mentioned in the introduction, the  ``$\nu$-deformation'' spoils the
integrability.
Remarkably that there exists  an integrable deformation of the BL model with $\nu=0$.
The corresponding 
model was introduced  by Fateev in the works
\cite{Fateev:1995ht,Fateev:1996ea} and it will  be referred to bellow 
as the Fateev  model.

Contrary to the BL model, the Fateev (F)  model involves three Bose fields 
governed by the Lagrangian
\bea\label{aposoasio}
{\tilde {\cal L}}_{\rm F}&=& \frac{1}{16\pi}\ \sum_{i=1}^3
\big(\, (\partial_0\varphi_i)^2-(\partial_1\varphi_i)^2\,\big)\\
&+&2\mu\
\big(\, \re^{\ri\, \alpha_3\varphi_3}\ \cos(\alpha_1\varphi_1+\alpha_2\varphi_2)+\re^{-\ri \alpha_3\varphi_3}\
\cos(\alpha_1\varphi_1-\alpha_2\varphi_2)\,\big)\,  .
\nonumber
\eea
Here $\alpha_i=\frac{1}{2}\sqrt{a_i}$ and
the    coupling constants $a_i$ satisfy   a single constraint
\bea\label{aposapoas}
a_1+a_2+a_3=2\ ,
\eea
which implies that the parameter $\mu$ has a dimension of mass.
As $\alpha_3\to 0$, the field $\varphi_3$ decouples 
in  \eqref{aposoasio} and
the interacting part 
coincides with the bosonic version of the BL Lagrangian \eqref{jssksjsak} with  $a_1+a_2=2$.
In fact, this  observation  requires a more careful assessment.
Performing the limit $\alpha_3\to 0$, 
one should expand the exponentials $\re^{\pm\ri\, \alpha_3\varphi_3}$ in \eqref{aposoasio} to the terms 
$\propto a_3=4 \alpha^2_3$. 
The  mass of the decoupled field is given by the relation 
$m^2=8\pi\mu a_3\, \langle\, \cos(\alpha_1\varphi_1)\cos(\alpha_2\varphi_2)\,\rangle$,
where the vacuum expectation value is taken for
the BL model with $\nu=0$.  This expectation value is simply related to the
corresponding specific bulk energy, 
$\langle\, \cos(\alpha_1\varphi_1)\cos(\alpha_2\varphi_2)\,\rangle=
-\frac{1}{4}\ \frac{\partial {\cal E}}{\partial \mu}$, and hence
\bea
\label{ossiis}
m^2=-2\pi \mu\, \lim_{a_3\to 0} \,\Big( a_3\ \frac{\partial {\cal E}}{\partial \mu}\,\Big)\ .
\eea
Eq.\,\eqref{oaisaiasoso}
shows that ${\cal E}=4\pi\mu^2\,\log(\mu\epsilon)+\ldots$ and, as  has been argued above,    should  be replaced by
${\cal E}=
\frac{4\pi\mu^2}{a_1+a_2-2}+\ldots$ within the analytical regularization. This, combined
with \eqref{ossiis} and  the constraint  $a_3=2-a_1-a_2$, 
means   that the field $\varphi_3$ has  the mass $m=4\pi\mu$ in the
decoupling limit. Taking into account $M-\mu$ relation \eqref{jshsyusy}, one finally obtains
\bea
m=
2M\, \cos\big({\textstyle\frac{\pi\delta}{2}}\big)\ ,
\eea
where we use $\delta=1-a_1=a_2-1$.

One of Fateev's  important results concerning the theory \eqref{aposoasio}
is an elegant analytical expression for the specific bulk energy \cite{Fateev:1996ea}:
\bea\label{akjsaskj}
{\cal E}_{\rm F}=-\pi\mu^2\ \prod_{i=1}^3\frac{\Gamma(\frac{a_i}{2})}{\Gamma(1-\frac{a_i}{2})}\ .
\eea
The linear constraint imposed on parameters $a_i$, can be resolved by setting
$a_1=1-\delta-\frac{a_3}{2}$ and $a_2=1+\delta-\frac{a_3}{2}$, and,  therefore, as $a_3\to 0$ one has
\bea\label{saoasiosiosao}
{\cal E}_{\rm F}=\pi \mu^2\, \Big( -\frac{2}{a_3}-4\log 2+
\psi\big({\textstyle\frac{1+\delta}{2}}\big)+\psi\big({\textstyle\frac{1-\delta}{2}}\big)-
2\psi\big({\textstyle\frac{1}{2}}\big)+o(1)\,\Big)\ .
\eea
Keeping in mind that $\frac{1}{a_3}$ can be substituted by
$(-\log(\mu\epsilon))$ one find
the relation
\bea
{\cal E}_{\rm F}\to {\cal E}+const\ m^2\ \ \ \ {\rm as}\ \ \ \ a_3\to 0\ ,
\eea
where ${\cal E}$ is  the specific  bulk energy 
for the BL model \eqref{oaisaiasoso},
whereas the term $\propto m^2$ 
is a  contribution of the free massive  field.
Notice that  $const$ does not depend on the coupling $\delta$, and 
it can be always set to  zero.

We can consider now  the Fateev model   in finite volume  with
the periodic boundary conditions $\varphi_i(x^0,x^1+R)= \varphi_i(x^0,x^1)$ imposed on all three fields $\varphi_i$.
Similar to the definition  \eqref{osaoasisoais} for
the BL model,  let us introduce ${\mathfrak F}_{\rm F}=R\, ( {E}_{\bf k}-R{\cal E}_{\rm F})/\pi$.
Then the above  consideration  suggests that
\bea\label{apossoasop}
\lim_{a_3\to 0^-}{\mathfrak F}_{\rm F}={\mathfrak F}(r,{\bf k})+ 
{\mathfrak f}_{\rm B}\big(2r c\big({\textstyle\frac{\delta}{2}}\big)\big)\ ,
\eea
where the second term in the r.h.s. with
\bea
{\mathfrak f}_{\rm B}(\beta)=\frac{\beta}{2\pi^2}\ 
\int_{-\infty}^\infty\rd\theta\ \cosh(\theta)\ \log\big(1-\re^{-\beta\cosh(\theta)}\big)\ ,
\eea
corresponds to  a contribution of the free  boson of mass 
$2M c\big({\textstyle\frac{\delta}{2}}\big)$ with $c(x)\equiv\cos(\pi x)$. 
Notice that the limit in \eqref{apossoasop} should be taken from negative values of  $a_3$, so that
the Lagrangian \eqref{aposoasio}  is real.
For $a_i>0 \ (i=1,2,3)$ the Lagrangian is complex, but
the QFT  is still well defined. 
In this case the potential term in  \eqref{aposoasio}  
is periodic  w.r.t.  all  fields $\varphi_i$ and
the space of states splits on the
orthogonal
subspaces
characterized by a triple of quasimomenta  ${\bf k}=(k_1,k_2,k_3)$.
For $a_3<0$ different sectors of the theory    are labeled by a
pair of  quasimomenta, similar to the case of the BL model, so that
eq.\,\eqref{apossoasop} can be understood literally as  a relation between
the vacuum energies in the Fateev and BL models
characterized by the same  ${\bf k}=(k_1,k_2)$.

A major  advantage of the case with all positive $a_i$ is that
the general  structure of the small-$R$  expansion in this regime  is considerably simple compared to 
the case $a_3<0$.
For $a_i>0$ the potential term in the Lagrangian \eqref{aposoasio}
is a uniformly bounded perturbation for any finite value of 
the dimensionless product  $\mu R$. Therefore
the conformal perturbation theory
yields an expansion of the form
\bea\label{asopisspaasop}
\frac{R}{\pi}\, E_{\rm F}=-\sum_{n=0}^\infty e^{(\rm F)}_{2n}\ (\mu R)^{4n}\ \ \ \ \ 
\ \ (a_i>0)\ .
\eea 
Here $e^{(\rm F)}_{0}=\frac{1}{6}\sum_{i=1}^3\big(1-6\,a_i\,k_i^2\,\big)$, whereas
the coefficients $e^{(\rm F)}_{2n}$ for $n\geq 1$  are  expressed in terms of convergent 2D Coulomb-type integrals,
for example
\bea\label{asapasosapp}
e^{(\rm F)}_2&=&2
\int\prod_{i=1}^3\frac{\rd^2 z_i}{2\pi}\
|z_1|^{-1+2 p_1+2 p_2+2p_3 }\
|z_2|^{-1-2 p_1+2 p_2-2p_3}\
|z_3|^{-1-2p_1-2p_2+2p_3}\
\\
&\times& \big|(z_1-1)(z_2-z_3)\big|^{2 a_1-2}\
\big|(z_1-z_2)(z_3-1)\big|^{2 a_2-2}\
\big|(z_1-z_3)(z_2-1)\big|^{2 a_3-2}\ ,\nonumber
\eea
where $p_i=\frac{1}{2}\, a_ik_i$.
Notice that the integral diverges at  $a_3\to 0^+$ and  formula \,\eqref{aosiosoaaa}   for the
asymptotic coefficient $e_2(\delta)$ in the  BL model  is a regularized version
of  $e^{(\rm F)}_2$ with $p_3=0$. Similarly to the expression for 
$e_2(\delta)$, eq.\,\eqref{asapasosapp} can be brought to the form
\bea
\label{sopaps}
e^{(\rm F)}_2=\frac{1}{4\pi}\ \int\rd^2\zeta\
|\zeta|^{2 a_1-4}|1-\zeta|^{2 a_2-4}\ \big(\tau_{p_1 p_2 p_3}(\zeta)\big)^2\ ,
\eea
where
\bea\label{oaissisaoi}
\tau_{p_1p_2p_3}(\zeta)=
-\frac{1}{2p_1}\sum_{\sigma=\pm} \sigma\,  \sqrt{\Omega(\sigma p_1,p_2+p_3) \Omega(\sigma p_1,p_2-p_3)}
\ \ \big|\chi_{\sigma p_1,p_2,p_3}(\zeta)\big|^2
\eea
and
\bea
\chi_{p_1p_2p_3}(\zeta)=
\zeta ^{\frac{1}{2}+p_1}(1-\zeta)^{\frac{1}{2}+p_2}\
\, {}_2F_1\big({\textstyle \frac{1}{2}}+ p_1+p_2+p_3,
{\textstyle \frac{1}{2}}+ p_1+p_2-p_3, 1+2 p_1;\zeta\big)\ .
\eea
The derivation follows the same steps outlined in Sec.\,\ref{sec2};
Fist of all,  one should 
substitute the integration variables $z_2$
by $\zeta=\frac{(1-z_1)(z_2-z_3)}{(1-z_2)(z_1-z_3)}$.
Then the integral  over  $z_1$  is performed using the  identity \eqref{opasssaaspsp}
where $p_1$ is substituted by $p_1+p_3$.
Finally one should  use the identity generalizing \eqref{ahssaajh}:
\bea
\big(\tau_{p_1 p_2 p_3}(\zeta)\big)^2=
|\zeta|^2\ \int
\frac{\rd^2 z}{\pi|z|^2}\
\Big|\frac{1-\zeta z }{z (z-\zeta)}\Big|^{2p_1+2p_3}\ |z|^{4p_3}\
\frac{\tau_{p_1+p_3, p_2,0}\big(X(z)\big)}{|X(z)|}\ ,
\eea
where $X(z)=\frac{(\zeta-z)(1-\zeta z)}{\zeta(1-z)^2}$.
An important observation is that $\tau_{p_1p_2p_3}(\zeta)$,
considered as a function on the Riemann sphere,
is regular except   for three points  $\zeta=0,\,1,\infty$
and
\bea\label{liouville1}
\eta_{\rm L}= -\log \tau_{p_1p_2p_3}(\zeta)
\eea
is a real solution of the Liouville equation
\bea\label{liouville2}
\partial_\zeta\partial_{\bar \zeta}\eta_{\rm L}-\re^{2\eta_{\rm L}}=0
\eea
for  $|p_i|<\frac{1}{2},\ \sum_i |p_i|<\frac{1}{2}$
(for details, see e.g. ref.\,\cite{Zamolodchikov:1995aa}).  Notice that 
$\tau_{p_1p_2p_3}(\zeta)=\tau_{p_1p_2p_3}(1-\zeta)=|\zeta|^2\,\tau_{p_3p_2p_1}(\zeta^{-1})$ and
therefore 
$\eta_{\rm L}$  satisfy the following asymptotic conditions at the punctures:
\bea\label{paspapsoasa}
\eta_{\rm L}\to
\begin{cases}
(2|p_1|-1)\, \log|\zeta|+O(1)\ \ \ &{\rm as}\ \ \ \zeta\to 0\\ 
(2|p_2|-1)\, \log|\zeta-1|+O(1)\ \ \ &{\rm as}\ \ \ \zeta\to 1\\
(2| p_3|+1)\,  \log|1/\zeta|+O(1)\ \ \ &{\rm as}\ \ \ \zeta\to\infty
\end{cases}
\ .
\eea
This way the result of  conformal perturbation theory
can be expressed in terms of  solution of the Liouville equation on the 
three-punctured sphere $ {\mathbb S}^2/\{0,\,1,\infty\}$:
\bea\label{asopisspaop}
\frac{R}{\pi}\ E_{\rm F}=-\frac{1}{6}\ \sum_{i=1}^3\Big(1-\frac{24}{a_i}\ p_i^2\,\Big)-
\frac{1}{4\pi}\ \int\rd^2\zeta\  |P(\zeta)|^2\ \re^{-2\eta_{\rm
    L}}+O(\rho^8)\ \ \ \ (a_i>0)\ , 
\eea
where  $\rho=\frac{1}{2}\ \mu R$ and
\bea
P(\zeta)=\rho^2\ \zeta^{ a_1-2}(1-\zeta)^{ a_2-2}\ .\label{Pzeta-def}
\eea

In ref.\,\cite{Lukyanov:2013wra} it was conjectured that
\bea\label{asopisspaopsdd}
\frac{R}{\pi}\ E_{\rm F}=-\frac{1}{6}\ \sum_{i=1}^3\Big(1-\frac{24}{a_i}\ p_i^2\,\Big)-
\frac{1}{4\pi}\ \int\rd^2\zeta\  |P(\zeta)|^2\ \re^{-2\eta}\ \ \ \ \ \ \ (a_i>0)\ ,
\eea
where $\eta$ is a real  solution of the so-called  modified sinh-Gordon equation
\bea\label{aopsasapso}
\partial_\zeta\partial_{\bar \zeta}\eta-\re^{2\eta}+|P(\zeta)|^2\ \re^{-2\eta}=0\ ,
\eea
satisfying the 
the same  asymptotic conditions as \eqref{paspapsoasa} (i.e., $\eta_{\rm L}$ should be 
substituted by $\eta$ in \eqref{paspapsoasa}). 
The  last term in \eqref{aopsasapso} $\propto \rho^4$ and   can 
be treated  perturbatively for   $|p_i|<\frac{a_i}{4}$.
Therefore  the small-$R$ behavior \eqref{asopisspaop}  follows immediately from  the exact
formula \eqref{asopisspaopsdd}.
One can show  that  the leading large-$R$ asymptotic
of \eqref{asopisspaopsdd} correctly reproduces  the specific bulk energy \eqref{akjsaskj} (see ref.\cite{Lukyanov:2013wra} for details).
Additional  arguments in  support  of eq.\,\eqref{asopisspaopsdd}  were presented  in the work \cite{Bazhanov:2013cua}.

\begin{figure}
\centering
\psfrag{a}{$\pi a_1$}
\psfrag{b}{$\pi a_2$}
\psfrag{c}{$\frac{\pi a_1}{2}$}
\psfrag{d}{$\frac{\pi a_2}{2}$}
\psfrag{e}{$\frac{\pi a_3}{2}$}
\psfrag{w1}{$w_1$}
\psfrag{w2}{$w_2$}
\psfrag{w3}{$w_3$}
\psfrag{bw3}{${\bar w}_3$}
\includegraphics[width=4.5cm]{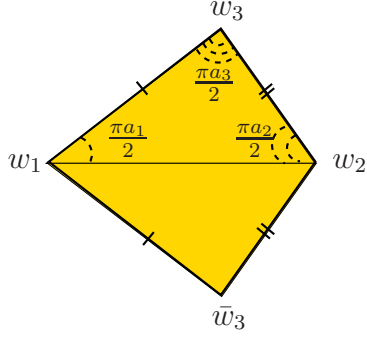}
\caption{Triangle $(w_1, w_2, w_3)$ is a $w$-image of the 
upper half plane $\Im m(\zeta)>0$ under the  Schwarz-Christoffel mapping \eqref{soiosa} with $a_i>0$. 
The point ${\bar w}_3$ is a reflection of $ w_3$ w.r.t. the straight line $(w_1,w_2)$. The domain  ${\mathbb D}_{\rm F}^{(+)}$ is obtained
from the 4-polygon  $(w_1, w_3, w_2,{\bar w}_3)$
by  the  identification
of the sides $[w_1,w_3]\sim[w_1,{\bar w}_3]$ and $[w_2,w_3]\sim [w_2,{\bar w}_3]$.}
\label{fig8}
\end{figure}

\begin{figure} 
\centering
\psfrag{a}{$\pi a_1$}
\psfrag{b}{$\pi a_2$}
\psfrag{c}{$\frac{\pi a_1}{2}$}
\psfrag{d}{$\frac{\pi a_2}{2}$}
\psfrag{e}{$\frac{\pi a_3}{2}$}
\psfrag{w1}{$w_1$}
\psfrag{w2}{$w_2$}
\psfrag{w3}{$w_3$}
\psfrag{bw3}{${\bar w}_3$}
\includegraphics[width=4cm]{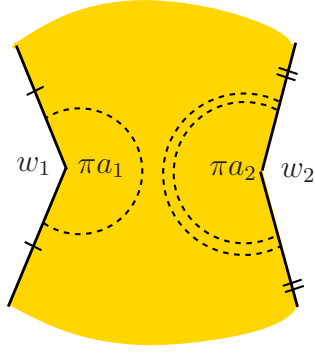}
\caption{Domain  ${\mathbb D}_{\rm F}^{(-)}$ --
the image of the 
thrice-punctured sphere for the case of    Schwarz-Christoffel mapping \eqref{soiosa} with $a_{1,2}>0$,
  $a_3<0$.}
\label{fig3}
\end{figure}

\begin{figure}
\centering
\psfrag{a}{$\pi a_1$}
\psfrag{b}{$\pi a_2$}
\psfrag{c}{$r/4$}
\psfrag{w1}{$w_1$}
\psfrag{w2}{$w_2$}
\psfrag{w}{$w$}
\includegraphics[width=4.5  cm]{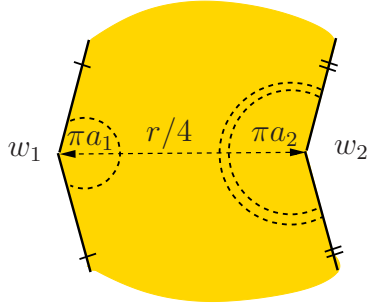}
\caption{Domain  ${\mathbb D}_{\rm BL}$ --
the image of the 
thrice-punctured sphere for the case of    Schwarz-Christoffel mapping \eqref{soiosa} with
$a_1+a_2=2$. The overall size of  ${\mathbb D}_{\rm BL}$ is controlled by
a length of the segment $(w_1, w_2)$,  which coincides with $r/4$}
\label{fig3sgsg}
\end{figure}

Eq.\,\eqref{asopisspaopsdd}   can be transformed to a formula
for the scaling function ${\mathfrak F}_{\rm F}\equiv R\,(E_{\rm F}-R{\cal E}_{\rm F})/\pi$. For this purpose, one  should consider 
the  Schwarz-Christoffel mapping
\bea\label{soiosa}
w(\zeta)=\int\rd \zeta\ \sqrt{P(\zeta)}\ ,
\eea
which maps the upper half plane $\Im  m(\zeta)\geq 0$ to the triangle $(w_1,\,w_2,\,w_3)$ in the complex $w$-plane
(see Fig.\,\ref{fig8}).
The lower half plane $\Im  m(\zeta)\leq 0$ is   mapped into the congruent triangle   $(w_1,w_2,{\bar w}_3)$.
It is straightforward  to show that the real function 
${\hat \eta}={ \eta}-{\textstyle \frac{1}{4}}\ \log(P{\bar P})$
is a solution of the  sinh-Gordon equation
\bea\label{sinh-eq}
\partial_w\partial_{\bar w}{\hat \eta}-\re^{2{ \hat \eta}}+ \re^{-2{ \hat \eta}}=0
\eea
in the open  domain  ${\mathbb D}_{\rm F}^{(+)}$,  which is  obtained  by gluing together
the triangles along
their    sides,   as it shown in Fig.\,\ref{fig8}. At the singular
points $w=w_i\ (i=1,2,3)$
the solution has
the following  asymptotic behavior:
\bea\label{osasail}
{\hat\eta}=(2 |k_i|-1)\ \log|w-w_i|+O(1)\ \ \ \ \ \ \ \ {\rm as}\ \ \ \ \ w\to w_i\  .
\eea
In ref.\,\cite{Lukyanov:2013wra} it was shown that  formula \eqref{asopisspaopsdd}  implies the relation 
 \bea\label{soapsopsa}
{\mathfrak F}_{\rm F}=-{ \frac{8}{\pi}}\  \int_{{\mathbb D}_{\rm F}^{(+)}}\rd^2 w\  \sinh^2({\hat\eta})+
\sum_{i=1}^3  a_i \, \big(|k_i|-{\textstyle\frac {1}{2}}\big)^2\ \ \ \ \ \ (a_i>0)\ .
\eea
Then, in  the consequent paper \cite{Bazhanov:2014joa}, it was argued that   \eqref {soapsopsa},
 with  some  minor modifications, also applies   to the case $a_{1,2}>0$, $a_3<0$.
Namely,
\bea\label{soapsopsadddf}
{\mathfrak F}_{\rm F}=-{ \frac{8}{\pi}}\  \int_{{\mathbb D}^{(-)}_{\rm F}}\rd^2 w\  \sinh^2({\hat\eta})+
\sum_{i=1}^2  a_i\, \big(|k_i|-{\textstyle\frac {1}{2}}\big)^2 \  \ \ \ \  \  (a_1,a_2>0,\ a_3<0)\ ,
\eea
where now ${\hat \eta}$ is  a solution of the  sinh-Gordon equation
\eqref{sinh-eq} in the domain shown in Fig.\,\ref{fig3},
satisfying  the asymptotic conditions \eqref{osasail}  at the vertices $w_1$ and $w_2$,
and
\bea
{\hat \eta}\to 0\ \ \ \ \ {\rm as}\ \ \ \ |w|\to\infty\ .
\eea
As $a_3\to 0^-$,  the domain ${\mathbb D}_{\rm F}^{(-)}$  tends to the region ${\mathbb D}_{\rm BL}$ 
shown in   Fig.\,\ref{fig3sgsg}.
With the relation \eqref{apossoasop}, this leads to  the following
exact formula for the scaling function
${\mathfrak F}(r,{\bf k})$ in the BL model,
\bea\label{soapsopsaddf}
{\mathfrak F}(r,{\bf k})=-{\mathfrak f}_{\rm B}\big(2r c\big({\textstyle\frac{\delta}{2}}\big)\big)
-{ \frac{8}{\pi}}\  \int_{{\mathbb D}_{\rm BL}}\rd^2 w\  \sinh^2({\hat\eta})+
\sum_{i=1}^2  a_i\, \big(|k_i|-{\textstyle\frac {1}{2}}\big)^2  \ .
\eea
As we shall see below this formula is in a perfect agreement with all
perturbation theory calculations, considered in Sec.\,\ref{sec2} and
Sec.\,\ref{sec3} of this paper, as well as with all other known 
results on the BL model, including the Bethe ansatz results of 
\cite{Bukhvostov:1980sn,Saleur:1998wa}.


The sinh-Gordon equation \eqref{sinh-eq} is 
a classical integrable equation which 
can be treated by the inverse scattering transform method. 
Thus the relation \eqref{soapsopsaddf} allows one to  apply this
powerful method  to  the problem of determining 
the vacuum energies. The working is very similar to that for the
Fateev model, considered in
\cite{Bazhanov:2013cua}, where all $a_1,a_2,a_3>0$, though contains
a few original details.
We postpone these derivations to our future publication \cite{BLR:2016b}
but present the final result here.
The scaling function \eqref{soapsopsaddf} is expressed through
the solution of a system of two Non-Linear Integral Equations (NLIE):
\bea\label{DDV}
\varepsilon_\sigma(\theta)=r\,\sinh(\theta-\ri\chi_\sigma)-2\pi k_\sigma+
\sum_{\sigma'=\pm}
\int_{-\infty}^{\infty}\frac{\rd\theta'}{\pi}\,
{G}_{\sigma\sigma'}(\theta-\theta')\
\Im m\Big[\log\big(1+e^{-\ri\varepsilon_{\sigma'}(\theta'-\ri 0)}\big)\Big].
\eea
Here $\sigma=\pm$,  $(\chi_+,\chi_-)=(0,\,\pi a_1/2)$
and the kernels are given by the relations
\bea\label{kerdef}
G_{\pm\pm}(\theta)=G_{a_1}(\theta)+G_{a_2}(\theta)\ ,\ \ \ \ \ \ 
{ G}_{\pm\mp}(\theta)={\hat G}_{a_1}(\theta)-{\hat G}_{a_2}(\theta)\ .
\eea
with
\bea
G_a(\theta)&=&
\int_{-\infty}^\infty\rd\nu \ \frac{\re^{\ri\nu\theta}\, \sinh(\frac{\pi\nu}{2}(1-a))}
{2\cosh(\frac{\pi \nu}{2})\sinh(\frac{\pi \nu a}{2})}\\
{\hat G}_a(\theta)&=&
\int_{-\infty}^\infty \rd\nu\ \frac{\re^{\ri\nu\theta}\, \sinh(\frac{\pi\nu }{2})}
{2\cosh(\frac{\pi \nu}{2})\sinh(\frac{\pi \nu a}{2})}\ .\nonumber
\eea
Once the numerical data for $\varepsilon_\pm(\theta)$ are available, ${\mathfrak F}(r,{\bf k})$ \eqref{soapsopsaddf}
can be computed by means of the relation
\bea\label{aooisaosa}
{\mathfrak F}(r,{\bf k})=\pm \frac{ r}{\pi}\ \Im m\Big[L_+(\pm \ri)+
\re^{\mp\frac{\ri\pi}{2}a_1}\,L_-(\pm\ri)\,
\Big]\ ,
\eea
where
\beq\label{kajsjas}
L_\sigma(\nu)=\int_{-\infty}^\infty\frac{\rd \theta}{\pi}\  \re^{-\ri \nu \theta}\,
\log\big(1+e^{-\ri\varepsilon_\sigma(\theta-\ri 0)}\big)\ .
\eeq
Notice that   \eqref{aooisaosa} is  valid for  both choices  of  the
sign $\pm$. 

\begin{figure}
\centering
\includegraphics[height=7.5cm]{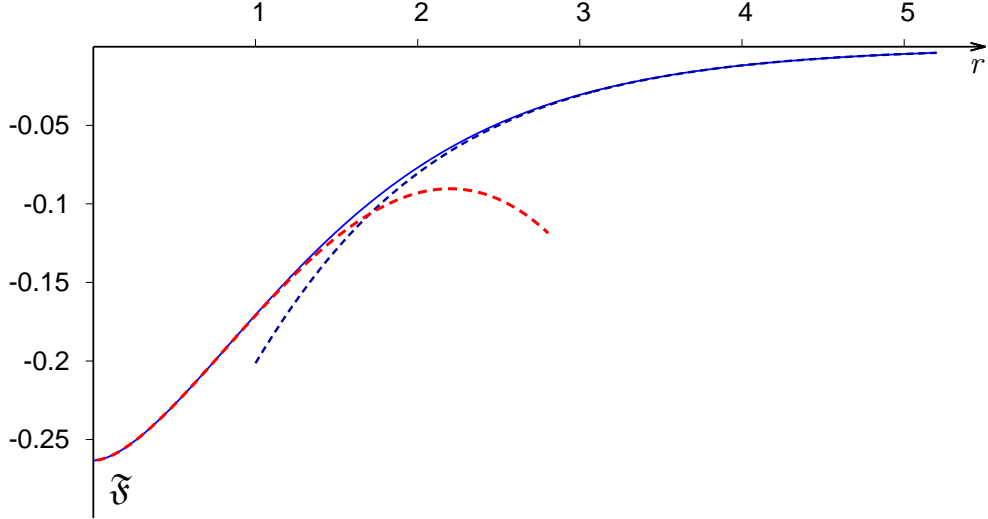}
\caption{
The scaling function ${\mathfrak F}(r,{\bf k})$ as a function of $r=
MR$ for
$\delta=\frac{17}{47}=0.36\ldots,\ k_1=\frac{47}{150},\ k_2=\frac{47}{640}$. 
The solid line was obtained from numerical
integration of  \eqref{DDV},\,\eqref{aooisaosa}.
The blue dashed  and red  dotted  lines represent,  respectively, 
the large-$r$  approximation\ \eqref{hauasyu}
and the small-$r$ expansion \eqref{apspassp}. For the chosen set of
parameters the latter becomes
${\mathfrak F}=-0.263322916666667-0.0719853960038915\, r^2\,\log(r)+0.092255549888030\, r^2+
0.0000477491676\, r^4+O(r^6)$\,. The numerical values
for ${\mathfrak F}$ and its asymptotics are given in Tab.~\ref{tab1} on
page~\pageref{tab1}. 
}
\label{fig5bb}
\end{figure}

Eq.\eqref{aooisaosa} can  be compared against the predictions of
renormalized perturbation theory in several ways. First, note that the
integral equation \eqref{DDV} have a  smooth limit for $\delta\to0$ (its
kernel vanishes linearly in $\delta$). 
Using this property we have verified that the function ${\mathfrak
  F}_2$ in eq.\,\eqref{apospapaso}, extracted from the 
numerical solution of \eqref{DDV}-\eqref{kajsjas}  for $k_1=k_2=0$ and
 $0.1\leq r\leq 5$, within nine significant digits coincides 
with the result of the perturbative calculations, shown with the solid
line in the right panel of Fig.\,\ref{fig4}.
Second, one can show that the exact formula \eqref{aooisaosa} implies the
following large-$R$ asymptotics
%
%
\bea\label{hauasyu}
&&{\mathfrak F}(r,{\bf k})={\mathfrak F}_{0}(r,{\bf k})+
{\mathfrak f}_{ \rm B}(2r)-{\mathfrak f}_{\rm B}\big(2r c\big({\textstyle\frac{\delta}{2}}\big)\big)
+
\frac{16 r}{\pi^2}\ \sum_{i=1}^2\int_{-\infty}^\infty \frac{\rd\nu}{2\pi}\
\\
&&\times\ 
 \Big(c^2(k_1)\,c^2(k_2)-c^2(k_{3-i})\,\cosh^2\big({\textstyle \frac{\pi\nu}{2}}\big) \Big)\ 
K_{\ri\nu}(r)K_{1-\ri\nu}(r)
\ \frac{\sinh(\frac{\pi\nu}{2}(1-a_i))}
{\cosh(\frac{\pi\nu}{2})\sinh(\frac{\pi\nu}{2} a_i)}+o\big(\re^{-2r}\big)\ ,\nonumber
\eea
where $k_1=k_++k_-,\,k_2=k_+-k_-$ and $c(x)\equiv\cos(\pi x)$.
Expanding this relation to the second order in $\delta=1-a_1=a_2-1$, one finds that the result  is consistent with 
eqs.\,\eqref{apospapaso},\,\eqref{oaaaaasisao} and \eqref{sauiuias}
from Sec.\,\ref{sec3}. 
Third, the numerical values for ${\mathfrak
  F}(r,{\bf k})$ obtained from \eqref{aooisaosa} and 
presented in Fig.\,\ref{fig5bb} and Tab.\,\ref{tab1} on page~\pageref{tab1},  
show an excellent agreement with the large-$R$ asymptotic formula
\eqref{hauasyu} and also with the predictions of 
the conformal perturbation theory, given by \eqref{apspassp},
\eqref{aosiosaaso} and \eqref{aasopaps}. 

Finally, the exact expressions \eqref{soapsopsaddf} and \eqref{aooisaosa}
perfectly agree with the Bethe ansatz results, considered in the next section.

\section{Bethe ansatz results\label{sec5} }

As shown already in the original BL paper \cite{Bukhvostov:1980sn} the
fermionic model \eqref{Lagr1}  
could be solved by the coordinate Bethe ansatz. 
In this section we review and extend their results.   
Within the Bethe ansatz approach the
eigenvectors and eigenvalues of the Hamiltonian are parameterized through 
rapidities of pseudoparticles filling the bare vacuum state. 
These rapidities are determined by the Bethe Ansatz Equations (BAE). 
In the context of relativistic QFT models the number of pseudoparticles is
infinite and, therefore, the related BAE require
some regularization which makes that number finite. 
Following the BL paper \cite{Bukhvostov:1980sn} here we will 
impose a straightforward cutoff to the number of pseudoparticles. An
alternative and in many respects more efficient lattice-type regularization is
considered in our next paper \cite{BLR:2016b}. 

Let $N\ge2$ be an even integer. 
The BAE of ref.\cite{Bukhvostov:1980sn} involve two sets of unknown rapidities
(called Bethe roots) $\{u_\ell\}$ and $\{\theta_\JJ\}$,
containing $N$ and
$2N$ variables, where 
\beq
\ell \in\big\{\textstyle-\frac{N}{2}+1,-\frac{N}{2}+2,\ldots,\frac{N}{2}\big\}\,,
\qquad {\JJ}\in\big\{-N+1,-N+2,\ldots,N\big\}\,.\label{set2}
\eeq
Throughout this section we will assume that the indices $\ell$ and
${\JJ}$ always run over the above sets of values,  respectively.
With a slight change of notations  and some minor corrections\footnote{%
The parameter $g$ in \cite{Bukhvostov:1980sn} is related to our $\delta=-g$; their
integer $n$ is replaced here by $N$ (we assume that this number is even); 
we have restored a missing minus sign in the
LHS of eqs.\,(82) of \cite{Bukhvostov:1980sn}, which corresponds to ours
eq.\,\eqref{bae1}; 
the case of untwisted boundary conditions, considered
in \cite{Bukhvostov:1980sn},
corresponds to $p_1=p_2=0$ here.} the 
Bethe ansatz equations of ref.\cite{Bukhvostov:1980sn} (generalized
for the twisted boundary conditions \eqref{apssspps}) can be written as  
\begin{subequations}\label{bae}
\bea
-1&=&\ds\re^{2\pi\ri(p_1-p_2)}\,
\re^{\ri \Mba R\sinh
  \theta_\JJ}\ \prod_\ell\  
\frac{\sinh\big(\theta_\JJ-u_\ell-\hf\,\ri \pi \delta\big)} 
{\sinh\big(\theta_\JJ-u_\ell+\hf\,\ri \pi \delta\big)}\,\label{bae1}
\\[.8cm] 
-1&=&\ds\re^{-4\pi\ri p_1}\,\prod_{\ell'}
\frac{\sinh\big(u_\ell-u_{\ell'}+\ri \pi \delta\big)}
{\sinh\big(u_\ell-u_{\ell'}-\ri \pi \delta\big)}\ 
\prod_\JJ\  \frac{\sinh\big(u_\ell-\theta_\JJ-\hf\,\ri \pi \delta\big)}
{\sinh\big(u_\ell-\theta_\JJ+\hf\,\ri \pi \delta\big)}\,,\label{bae2}
\eea
\end{subequations}
where and 
the indices $\ell,\ell',\JJ$ take the integer values \eqref{set2}. 
The parameters $p_1$ and $p_2$ are defined by eqs.\,\eqref{apssspps}, \eqref{apssspps2} and \eqref{kasoisau}.
Altogether there are $3N$ equations for $3N$
unknown $\theta$'s and $u$'s.  
When the cutoff is removed, $N\to\infty$,\  the number of Bethe roots 
becomes  infinite. The parameter ${\cal M}$ is the {\em bare mass}
parameter entering the coordinate Bethe ansatz calculation of
\cite{Bukhvostov:1980sn} (denoted as ``$m$'' therein). Its
relationship with the physical fermion mass $M$ used in the previous
sections follows from the requirement that the scaling function, determined by the BAE, at  
large distances should vanish as $\propto\exp(-M R)$, i.e., 
exactly as the one in \eqref{aosioasiasoas}. As we shall see below
this  is achieved if one sets (see remarks after eq.\,\eqref{ceff-lat})  
\beq\label{physmass}
{\cal  
M}=M\cos\big({\textstyle\frac{\pi\delta}{2}}\big)\,.
\eeq
This relation will be assumed in what follows.
For practical purposes it is useful to rewrite BAE \eqref{bae} in
the logarithmic form
\begin{subequations}\label{baelog}
\bea
m_\JJ&=&{\hf}+p_1-p_2+\frac{\Mba R}
{2\pi}\ \sinh(\theta_\JJ)+ \sum_\ell\  {\mye}_{2\delta}(\theta_\JJ-u_\ell)
\label{baelog1}
\\[.8cm] 
{\overline m}_\ell&=&{\hf}-2p_1 -\sum_{\ell'}
{\mye}_{4\delta}(u_\ell-u_{\ell'})
+\sum_\JJ{\mye}_{2\delta}(u_\ell-\theta_\JJ)\,,\label{baelog2}
\eea
\end{subequations}
where 
\beq\label{e-def}
{\mye}_\alpha(\theta)=\frac{1}{2\pi\ri}\log\left[\frac{\sinh
\big(\frac{1}{4}\,
\ri \pi \alpha-\theta\big)}
{\sinh\big(\frac{1}{4}\,\ri \pi \alpha+\theta\big)}\right]\,,
\eeq
and the integer phases $\{m_\JJ\}$ 
and $\{{\overline m}_\ell\}$ play the r${\hat {\rm o}}$le of quantum numbers, which uniquely 
characterize solutions of the BAE.
Different solutions define different eigenstates
of the Hamiltonian. The energy of the corresponding state reads
\beq
E=-\Mba
\,\sum_\JJ \cosh({\theta_\JJ})\,. \label{e-lip}
\eeq
As usual, the most difficult question in the analysis of BAE is to 
determine patterns of zeroes and the corresponding phase assignment
in \eqref{baelog} for different states, in particular for the vacuum state.
For the untwisted boundary conditions, $p_1=p_2=0$, this question was
studied in \cite{Bukhvostov:1980sn}. It was shown that for small values of 
$|\delta|\ll 1$ the vacuum roots $\{u_\ell\}$ are real and
their positions are given by an asymptotic formula 
\beq\label{uass}
\Mba R\,\sinh u_\ell=(2\ell-1)\pi\,  
+O(\delta)\,\qquad\ \ \  (|\delta|\ll 1)\,,
\eeq
whereas the roots $\{\theta_\JJ\}$ split into pairs  
\beq \label{pairs}
\theta_{2\ell-\hf\pm\frac{1}{2} }=u_\ell\pm\sqrt{\frac{\pi\delta}{r
    \cosh( u_\ell)}}+O(\delta^{\frac{3}{2}})\,,
\eeq
centered around $u$'s. This description is valid for both signs of delta.
For $\delta>0$ the $\theta$-roots
are real and the phases in \eqref{baelog} take consecutive integer
values
\beq\label{phases}
m_\JJ=\JJ\,,\qquad {\overline m}_\ell=\ell\ \qquad (\delta>0)\,,
\eeq
within the range defined in \eqref{set2}. 
For $\delta<0$ the $u$-roots remain real and retain the same phases as
in \eqref{phases},
\begin{subequations}\label{phases2}  
\beq\label{phases2a}
\qquad {\overline m}_\ell=\ell\ \qquad ( \delta<0)\,.
\eeq
The 
$\theta$-roots become complex and form the so-called 2-strings with
a more subtle phase assignment. Near the origin $\big|\Re e
(\theta_\JJ)\big|<2/(\pi^2\delta)$ the phases are 
still consecutive, as stated in \cite{Bukhvostov:1980sn}\footnote{%
The phases of complex roots are not uniquely defined. Here we adopt
the convention that the functions \eqref{e-def} entering
\eqref{baelog} should not have jumps under small variation of roots
near their exact positions. For that reason for $\delta<0$ we replace ${
  \mye}_{2\delta}(\theta)$ in  \eqref{baelog} with 
$\tilde{\mye}_{2\delta}(\theta)$, where
$$
\tilde{\mye}_\alpha(\theta) =\frac{1}{2\pi\ri}\, \log\left[\frac{\sinh
\big(\theta-\frac{1}{4}
\ri \pi \alpha\big)}
{\sinh\big(\theta+\frac{1}{4}\ri \pi \alpha\big)}\right]\,,
$$
differs from \eqref{e-def} by the sign of the argument of the logarithm.
As a result our 2-strings phases assignment in \eqref{phases2b} looks 
different, but nevertheless equivalent to the corresponding eq.\,(92) 
in \cite{Bukhvostov:1980sn}. }   
\beq\label{phases2b}
m_{2\ell-\hf\pm\frac{1}{2}} =-\ell+1
\ \qquad (\delta<0)\,,
\eeq
however for larger $\big|\Re e(\theta_\JJ)\big|$ this is no longer true and
the consecutive phase segments are divided by regions of ``holes'',
where the RHS of \eqref{phases2b} jumps over several integers. A
general description of this pattern is unknown. 
\end{subequations}

\begin{figure}
\centering
\hspace*{-1cm}
\includegraphics[width=16cm]{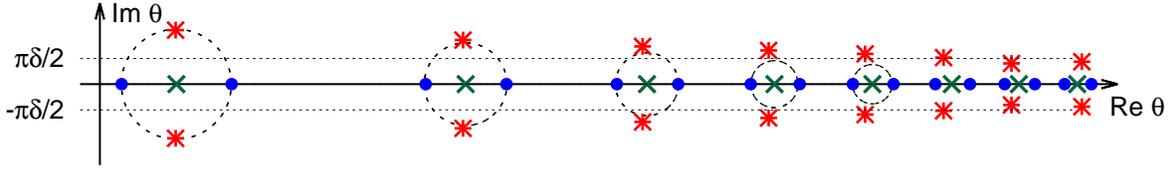}
\caption{The arrangements of the Bethe roots solving \eqref{baelog}
  with $N=16$, $p_1=p_2=0$ and $|\delta|=0.05$. The (green) crosses show
the roots $\{u_\ell\}$, (blue) dots show the roots $\{\theta_\JJ\}$
for $\delta>0$ and (red) asterisks show the (complex) roots
$\{\theta_\JJ\}$ for $\delta<0$ (the roots $\{u_\ell\}$ remains the
same). Only a part of complex plane, containing a half of the roots is
shown. The dashed lines and circles
illustrate the pairing of $\theta$-roots described by 
eqs.\,\eqref{pairs} and \eqref{pairs2}.}
\label{figsimple}
\end{figure}

The arguments of \cite{Bukhvostov:1980sn} are based on the perturbation theory around
the free fermion case with the untwisted boundary conditions 
(corresponding to $\delta=0$ and $p_1=p_2=0$) and expected to work well 
for sufficiently small $\delta$'s and vanishing $p$'s. We have
verified this picture numerically. The arrangement of the vacuum roots
for $N=16$ and $|\delta|=0.05$ is illustrated in Fig.\,\ref{figsimple},
where only a part of complex plane, containing a half of the roots is
shown.  For
$\delta<0$ the formula \eqref{pairs} is valid for
$|\ell|<2/(\pi^2\delta)$. For larger values of $\ell$ 
the $\theta$-roots form
almost perfect 2-strings
\beq \label{pairs2}
\theta_{2\ell-\hf\pm\frac{1}{2} }=u_\ell\pm  {\textstyle\frac{ 1}{2}}\, \ri\pi\delta
 \, \big(1+O(\ell^{-1})\big)\, ,\qquad \ell\gg2/(\pi^2|\delta|)\,\qquad
(\delta<0)\, . 
\eeq

Our numerical 
analysis shows that essentially the same picture of zeroes\footnote{%
When $p_1,p_2\not=0$, eqs.\,\eqref{uass} and \eqref{pairs} 
should be modified, but \eqref{phases}, \eqref{phases2} and
\eqref{pairs2} remain intact.}
 holds also for small non-zero
values of $p_1$ and $p_2$. In particular, the integer phases
\eqref{phases} and \eqref{phases2} 
remains the same, as they cannot
change under continuous deformations of the boundary conditions. 

\begin{figure}
\centering
\includegraphics[height=7.5cm]{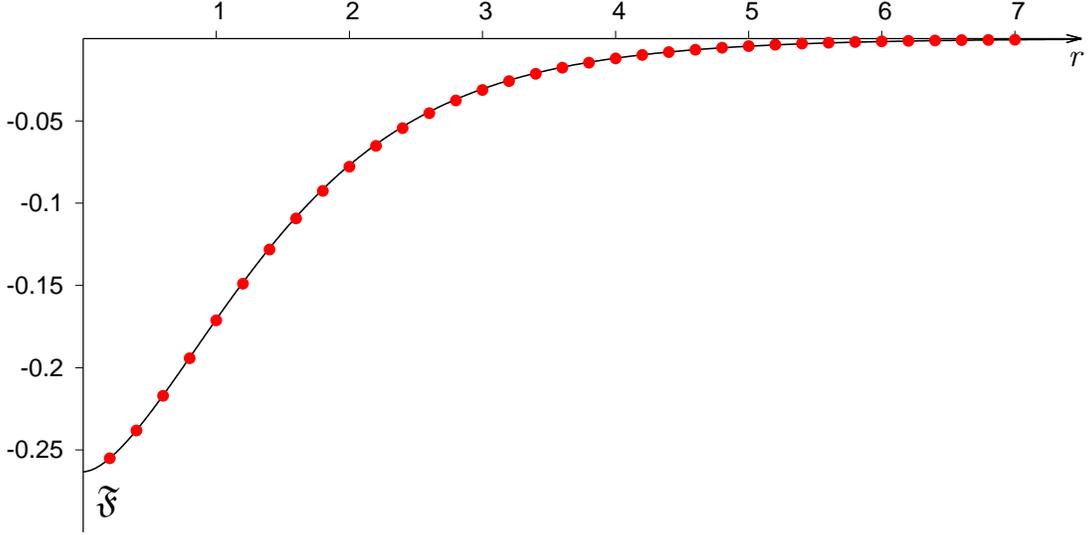}
\caption{Dots show values of ${\mathfrak F}(r,{\bf k})$ for
  $\delta=17/47$, $p_1=1/10,p_2=1/20$ calculated from \eqref{e-lip} and 
  \eqref{ceff-lat} with $N=500$ and the value of $C=0.9658605$.
 The continuous curve represents the results obtained from \eqref{aooisaosa}  and the NLIE \eqref{DDV}.} 
\label{ceff-pic}
\end{figure}

Using BAE \eqref{baelog} one can show \cite{BLR:2016b} that the vacuum energy
\eqref{e-lip} diverges quadratically for large $N$
(cf. eq.\,\eqref{paopsaoaops}) 
\beq
\frac{RE}{\pi}= \epsilon_{2}\, N^2
+{\epsilon}_0 \,r^2\,\log\big({4N/r}\big)+
O(1)\ \ \ \qquad (N\to \infty)\,,\label{qdiv}
\eeq
where 
\beq
\epsilon_2=-(1+\delta)\,,\qquad 
{\epsilon}_0=
-{\textstyle\frac{1}{\pi^2}}\ \cos^2\big(\textstyle{\frac{\pi\delta}{2}}\big)\,,\qquad r=M R\,.
\eeq
Then from the finite-size scaling arguments (applied in the context of
the Bethe ansatz regularization of massive field theory models \cite{Destri:1994bv,Lukyanov:2011wd}) one  
expects that for $N\to\infty$ \ 
the regularized expression for the energy 
\beq
{\mathfrak F}(r,{\bf k})=-\frac{c_{\bf{k}}}{6}+
\lim_{\scriptstyle{\begin{subarray}{c}
N\to\infty\\
r-\mbox{\scriptsize{fixed}}
\end{subarray}}}\left(\frac{RE }{\pi}
-\epsilon_2\, N^2-{\epsilon}_0\,
r^2\ \Big(\log({4N/r})+C\,\Big)\right)\,,
\label{ceff-lat} 
\eeq
where $c_{\bf{k}}=\sum_{i=1}^2\big(1-6 a_ik_i^2\big)$, reduces to the scaling function ${\mathfrak F}(r,{\bf k})$
for the integrable case of  the  QFT model  \eqref{Lagr1},\,\eqref {Lagr2}. 
The constant $C$ is non-universal, it is determined by the requirement ${\mathfrak F}(r,{\bf k})\to0$ as $r\to \infty$.
The relation \eqref{physmass} follows from the 
requirement that \eqref{ceff-lat} has the same 
  large distance decay exponent as in \eqref{aosioasiasoas}.
For $\delta>0$  the formula \eqref{ceff-lat} has been verified numerically.
The values ${\mathfrak F}(r,{\bf k})$ obtained from \eqref{e-lip} with
the solution of 
\eqref{baelog1}, \eqref{baelog2} and \eqref{phases} for $N=500$
display a good agreement (to within at least  three decimal places)
with the more accurate results 
obtained from the NLIE  \eqref{DDV}, 
see Fig.\,\ref{ceff-pic}. 

Finally note that, as shown by Saleur \cite{Saleur:1998wa},
the BAE \eqref{baelog1}, \eqref{baelog2}, describing 
the $\delta>0$ vacuum state \eqref{phases} filled by the
real $\theta$-roots,   
can be converted to a set of NLIE. 
After some minor corrections\footnote{%
In the untwisted case $k_1=k_2=0$ our eq.\,\eqref{sddv} 
is equivalent to eq.\,(7) of
\cite{Saleur:1998wa} where one should restore a missed factor $1/(2\pi)$ 
in front of the kernel $\Phi_{ij}$ therein; our eq.\,\eqref{senergy} is
equivalent to eq.\,(8) of \cite{Saleur:1998wa} where one should remove an 
extra factor $L$ in the LHS.} these NLIE 
(generalized for the twisted boundary conditions \eqref{apssspps}) can be
written as 
\beq
\tilde{\varepsilon}_j(\theta)={\tilde  r}_j\, \sinh(\theta) -2\pi {\tilde k}_j+ \sum_{l=1,2}\,
\int_{-\infty}^{\infty}\frac{\rd\theta'}{\pi}\ 
\widetilde{G}_{jl}(\theta-\theta')\  \Im m\Big[
\log\big(1+\re^{-\ri\tilde{\varepsilon}_l(\theta'-\ri 0)}\big)\,\Big]\, ,\label{sddv}
\eeq
where $j=1,2$, 
\beq 
{\tilde r}_1=2r\,\cos\big({\textstyle\frac{\pi\delta}{2}}\big)\,,\quad  {\tilde r}_2=r\,,\quad {\tilde k}_1= k_2\,,\quad
{\tilde k}_2=k_+\,,\quad k_\pm={\textstyle\frac{1}{2}}\, (k_1\pm k_2)\,.
\eeq 
The kernel $\widetilde{G}_{jl}$ reads  
\beq
\widetilde{G}_{11}(\theta)=\int_{-\infty}^\infty\,\rd\nu\ \frac{\re^{\ri
      \nu \theta}\,\sinh(\frac{\pi\nu a_1}{2})}   
{\sinh(\frac{\pi\nu a_2}{2})}\,,\qquad
\widetilde{G}_{22}(\theta)=\int_{-\infty}^\infty\,\rd\nu\ \frac{\re^{\ri
      \nu \theta}\,\sinh^2(\frac{\pi\nu \delta}{2})}   
{\sinh(\frac{\pi\nu a_1}{2})\sinh(\frac{\pi\nu a_2}{2})}\,,
\eeq
and 
\beq
\widetilde{G}_{12}(\theta)=\widetilde{G}_{21}(\theta)
=\int_{-\infty}^\infty\,\rd\nu\ \frac{\re^{\ri
      \nu \theta}\,\sinh(\frac{\pi\nu}{2})}   
{\sinh(\frac{\pi\nu a_2}{2})}\ .
\eeq
Note that $\widetilde{G}_{22}(\theta)$  coincides
with ${G}_{++}(\theta)$ defined in \eqref{kerdef}. 
With these notations the scaling function \eqref{ceff-lat} can be written
as 
\beq
{\mathfrak F}(r,{\bf k})
=\frac{1}{\pi^2} \sum_{j=1,2}\, {\tilde r}_j\, \int_{-\infty}^\infty\,
\rd \theta\, \sinh(\theta) \,\Im m\Big[\log\big(1+\re^{-\ri
  \tilde{\varepsilon}_j(\theta-\ri 0)}\big)\,\Big]\ .\label{senergy}
\eeq
%
\begin{table}[h]
\begin{center}
\begin{tabular}{| c || l| l | l|}
\hline \rule{0mm}{3.6mm}
$r=MR$&$\ \ \ \ \ \ \ {\mathfrak F}(r,{\bf k})$&
$\ \ \ \ \ {\mathfrak F}(r,{\bf k})_{\rm UV}$& 
$\ \ \ \ \ {\mathfrak F}(r,{\bf k})_{\rm IR}$\\
\hline
$0.1$& $-0.2607428309788$   &$-0.2607428313953$                      & $\ \ \ \ \ \ \ \ \star\star\star$\\
$0.2$& $-0.2549983506999$   &$-0.2549983772536$                      & $\ \ \ \ \ \ \ \ \star\star\star$\\
$0.3$& $-0.2472190685352$   &$-0.2472193690897$                      & $\ \ \ \ \ \ \ \ \star\star\star$\\
$0.4$& $-0.2380056043350$   &$-0.2380072781157$                      & $\ \ \ \ \ \ \ \ \star\star\star$\\
$0.5$& $-0.2277756139968$   &$-0.2277819263012$                      & $\ \ \ \ \ \ \ \ \star\star\star$\\
$0.6$& $-0.2168482299168$   &$-0.2168668158811$                      & $\ \ \ \ \ \ \ \ \star\star\star$\\
$0.7$& $-0.2054791982549$   &$-0.2055252929751$                      & $\ \ \ \ \ \ \ \ \star\star\star$\\
$0.8$& $-0.1938786835672$   &$-0.1939794374587$                      & $\ \ \ \ \ \ \ \ \star\star\star$\\
$0.9$& $-0.1822213607225$   &$-0.1824212140898$                      & $\ \ \ \ \ \ \ \ \star\star\star$\\
$1.0$& $-0.1706526112907$   &$-0.1710196176110$                      & $-0.2014349564662$\\
$1.1$& $-0.1592926194023$   &$-0.1599255303999$                      & $-0.1847842117398$\\
$1.2$& $-0.1482393124726$   &$-0.1492751769847$                      & $-0.1692047325819$\\
$1.3$& $-0.1375706804828$   &$-0.1391926693109$                      & $-0.1547042796504$\\
$1.4$& $-0.1273467826821$   &$-0.1297919365909$                      & $-0.1412670230341$\\
$1.5$& $-0.1176116153451$   &$-0.1211782235966$                      & $-0.1288603066874$\\
$1.6$& $-0.1083949276224$   &$-0.1134492772872$                      & $-0.1174399287778$\\
$1.7$& $-0.0997140161108$   &$-0.1066963026919$                      & $-0.1069542273502$\\
$1.8$& $-0.0915754934205$   &$-0.1010047442851$                      & $-0.0973472061921$\\
$1.9$& $-0.0839770063651$   &$-0.0964549329272$                      & $-0.0885608931075$\\
$2.0$& $-0.0769088715072$   &$-0.0931226275591$                      & $-0.0805370881568$\\
$2.4$& $-0.0535707439372$   &$\ \ \ \ \ \ \ \ \ \star\star\star$     & $-0.0549560607657$\\
$2.8$& $-0.0369408994984$   &$\ \ \ \ \ \ \ \ \ \star\star\star$     & $-0.0374531791217$\\
$3.2$& $-0.0253532059419$   &$\ \ \ \ \ \ \ \ \ \star\star\star$     & $-0.0255388456222$\\
$3.6$& $-0.0173705422556$   &$\ \ \ \ \ \ \ \ \ \star\star\star$     & $-0.0174369596054$\\
$4.0$& $-0.0118987032850$   &$\ \ \ \ \ \ \ \ \ \star\star\star$     & $-0.0119222709924$\\
$4.4$& $-0.0081537173228$   &$\ \ \ \ \ \ \ \ \ \star\star\star$     & $-0.0081620342332$\\
$4.8$& $-0.0055903459505$   &$\ \ \ \ \ \ \ \ \ \star\star\star$     & $-0.0055932695721$\\
$5.2$& $-0.0038344869615$   &$\ \ \ \ \ \ \ \ \ \star\star\star$     & $-0.0038355117142$\\
$5.6$& $-0.0026307619638$   &$\ \ \ \ \ \ \ \ \ \star\star\star$     & $-0.0026311203213$\\
$6.0$& $-0.0018049899423$   &$\ \ \ \ \ \ \ \ \ \star\star\star$     & $-0.0018051150200$\\
$6.4$& $-0.0012382494264$   &$\ \ \ \ \ \ \ \ \ \star\star\star$     & $-0.0012382930098$\\
$6.8$& $-0.0008492127577$   &$\ \ \ \ \ \ \ \ \ \star\star\star$     & $-0.0008492279219$\\
$7.2$& $-0.0005821697868$   &$\ \ \ \ \ \ \ \ \ \star\star\star$     & $-0.0005821750559$\\
$7.6$& $-0.0003989064681$   &$\ \ \ \ \ \ \ \ \ \star\star\star$     & $-0.0003989082967$\\
$8.0$& $-0.0002731844458$   &$\ \ \ \ \ \ \ \ \ \star\star\star$     & $-0.0002731850797$\\
$8.4$& $-0.0001869773822$   &$\ \ \ \ \ \ \ \ \ \star\star\star$     & $-0.0001869776017$\\
$8.8$& $-0.0001278976621$   &$\ \ \ \ \ \ \ \ \ \star\star\star$     & $-0.0001278977381$\\
$9.2$& $-0.0000874328116$   &$\ \ \ \ \ \ \ \ \ \star\star\star$     & $-0.0000874328378$\\
$9.6$& $-0.0000597346969$   &$\ \ \ \ \ \ \ \ \ \star\star\star$     & $-0.0000597347060$\\
$10.$& $-0.0000407872423$   &$\ \ \ \ \ \ \ \ \ \star\star\star$     & $-0.0000407872454$\\
\hline
\end{tabular}
\end{center}
\caption{Numerical data for Fig.\,\ref{fig5bb}.
The first column contains numerical values of ${\mathfrak F}(r,{\bf
  k})$ obtained by solving the NLIE
\eqref{DDV},\,\eqref{aooisaosa}
 for 
$\delta=\frac{17}{47}=0.36\ldots,\ k_1=\frac{47}{150},\ k_2=\frac{47}{640}$.
The second and third columns contain the short- and large-distance
asymptotics of ${\mathfrak F}(r,{\bf
  k})$, given by \eqref{apspassp} and \eqref{hauasyu}, respectively.
}
\label{tab1}
\end{table}
Note, that even though the equations \eqref{sddv} look totally different from
\eqref{DDV} the resulting expression \eqref{senergy} for the scaling function  
is, in fact, exactly equivalent to \eqref{aooisaosa}. A complete proof
of this equivalence is presented in our next paper \cite{BLR:2016b}. It
is also worth noting that from 
the point of view of numerical analysis the system \eqref{DDV}
displays a much faster convergence than \eqref{sddv} and, therefore,
requires lesser computational resources. Moreover, the
system \eqref{DDV} is well suited for small $\delta$ analysis, whereas
the eq.\,\eqref{sddv} becomes singular for $\delta\to0$ (the latter fact has
already been noted in \cite{Saleur:1998wa}, where
the NLIE \eqref{sddv} for the untwisted case $k_\pm=0$ were
originally derived).

\section{Conclusion}
The Bukhvostov-Lipatov (BL) model \cite{Bukhvostov:1980sn}
describes weakly interacting
instantons and anti-instantons in the $O(3)$ non-linear sigma model in
two dimensions. 
In this paper we have studied various
aspects of the BL model with twisted boundary conditions, 
using all well-established 
approaches to 2D massive integrable QFT, including the conformal perturbation
theory (Sec.\,\ref{sec2}), the standard renormalized
perturbation theory (Sec.\,\ref{sec3}) and the Bethe ansatz
(Sec.\,\ref{sec5}). Moreover, in Sec.\,\ref{secnew5} 
we have proposed an exact formula
\eqref{soapsopsaddf} for the vacuum energy of the model, expressing it
via a special solution of 
the sinh-Gordon equation \eqref{sinh-eq} 
in the domain ${\mathbb D}_{\rm BL}$ (see
Fig.\,\ref{fig3sgsg}). The required solution $\hat\eta(w)$ 
decays at $|w|\to\infty$
and obey the boundary conditions \eqref{osasail} at the singular points $w_1$
and $w_2$. The connection to the classically integrable sinh-Gordon
equation is rather powerful, since it 
allows one to obtain the non-linear integral equations
\eqref{DDV}, determining the vacuum energy in the form \eqref{aooisaosa}. 
We have shown that this formula perfectly matches all
our perturbation theory calculations as well as the previously known
coordinate Bethe ansatz 
results of Bukhvostov and Lipatov \cite{Bukhvostov:1980sn}, and Saleur
\cite{Saleur:1998wa}. The comparisons were done both analytically (where
possible) and numerically. Complete proofs and derivations of our
exact results are postponed into the
forthcoming publication \cite{BLR:2016b}.   
The main idea of that work is to connect the
functional equations for connection coefficients for the auxiliary
linear problem for the sinh-Gordon equation \eqref{sinh-eq} to the
Bethe ansatz equations \eqref{bae}, arising from the coordinate Bethe
ansatz \cite{Bukhvostov:1980sn}. This requires rather substantial
works involving the particle-hole transformation and
lattice-type regularization of the BAE, as well as some generalization
of arguments of ref.\cite{Bazhanov:2013cua}, devoted to the Fateev model.

Clearly, further study of the BL model is desirable. 
Indeed, almost all the considerations in this paper concerns the
weak coupling regime $0<\delta<1$. However, the
most interesting regime is the strong coupling regime $\delta>1$, 
where the BL model admits a dual description as the so-called 
sausage model \cite{Fateev:1992tk}. Interestingly, this model turns into the
$O(3)$ NLSM, in the limit $\delta\to\infty$. This suggests that the
instanton counting becomes exact in the strong coupling limit of the
BL model. We intend to address this problem in the future.

The description of the vacuum state energy 
of the BL model in terms of the classical
sinh-Gordon equation can be viewed as an instance of 
a remarkable, albeit unusual  
correspondence between 
{\em integrable quantum field theories}  
and 
{\em integrable classical field theories}  in two dimensions, which  
cannot be expected from the standard quantum--classical correspondence
principle. 
In the past two decades this topic has undergone various conceptual
developments, which can be traced through the works
\cite{DT99b,Bazhanov:1998wj,Suzuki:2000fc,Dorey:2006an,Feigin:2007mr,
Bazhanov:2003ni,Fioravanti:2004cz,Lukyanov:2010rn,Dorey:2012bx,
Lukyanov:2013wra,Bazhanov:2013cua,Masoero:2015lga,Ito:2015nla}.    
The commonly accepted mystery of this correspondence is slightly unveiled
by our conformal perturbation theory calculations in Sec.\,\ref{sec2} and
Sec.\,\ref{secnew5}. Indeed eqs.\,\eqref{aasopaps} and \eqref{asopisspaop},
expressing the vacuum energy in terms of the solutions
\eqref{oaissisaoi}, \eqref{liouville1} of the Liouville equation
\eqref{liouville2} arise as a direct result of calculations, without
any additional assumptions. It would be interesting to check whether
these calculations can be generalized to other integrable QFTs where
the correspondence to classical integrable equations is known.
More generally, it
would be very important  to better understand connections of the above correspondence
to mathematical structures arising in 4D gauge theories 
\cite{Gaiotto:2009hg,Nekrasov:2009rc,Litvinov:2013zda}, calculations of
amplitudes of high energy scattering
\cite{Alday:2009dv,Basso:2013vsa,Bartels:2008ce} and  dualities in
finite dimensional quantum-mechanical systems \cite{Mironov:2012ba}.

\section*{Acknowledgment}
The authors thank  G.V.~Dunne, L.D.~Faddeev, A.R.~Its, L.N.~Lipatov,
L.A.~Takhtajan, \newline
V.O.~Tarasov, A.M.~Tsvelick  and A.B.~Zamolodchikov
for their  interest to this work and useful remarks.  

\bigskip
\noindent
The research of SL is supported by the NSF under grant number
NSF-PHY-1404056. 

\providecommand{\href}[2]{#2}
\begingroup\raggedright
\endgroup

\end{document}